\documentclass[10pt]{article}
\usepackage[a4paper,margin=1in]{geometry}

\usepackage{jheppub}

\usepackage[T1]{fontenc}

\usepackage{graphicx}
\usepackage{amssymb}
\usepackage{amsmath}
\usepackage{dsfont}
\usepackage{setspace}
\usepackage{slashed}
\usepackage{pifont}
\usepackage[capitalise]{cleveref}
\usepackage{multirow}
\usepackage{makecell}
\usepackage{tabularx}
\newcolumntype{C}[1]{>{\hsize=#1\hsize\centering\arraybackslash}X}

\newcolumntype{Z}{r<{\hspace{3mm}}}

\renewcommand{\arraystretch}{1.4}

\def\cA{\mathcal{A}}
\def\cT{\mathcal{T}}
\def\cG{\mathcal{G}}
\def\cI{\mathcal{I}}
\def\cM{\mathcal{M}}
\def\cF{\mathcal{F}}
\def\cR{\mathcal{R}}
\def\cH{\mathcal{H}}
\def\cO{\mathcal{O}}

\def\eps{\epsilon}
\def\vareps{\varepsilon}
\DeclareMathOperator{\tr}{\rm tr}

\def\trm{\tr_-}
\def\trp{\tr_+}

\def\iamp{z}

\def\bbh{b\bar{b}H}
\def\bbggh{\bar{b}bggH}
\def\bbqqh{\bar{b}b\bar{q}qH}

\def\bbggH{0\to\bar{b}bggH}
\def\bbqqH{0\to\bar{b}b\bar{q}qH}
\def\bbbbH{0\to\bar{b}b\bar{b}bH}
\def\gg{\mathrm{gg}}
\def\qq{\mathrm{q\bar{q}}}

\def\la{\langle}
\def\ra{\rangle}
\def\spA#1#2{\la#1#2\ra}
\def\spB#1#2{[#1#2]}
\def\spAB#1#2#3{\la#1|#2|#3]}

\renewcommand{\i}{\ensuremath{\mathrm{i}}}

\def\colhelsum{\underset{\mathrm{colour}}{\overline{\sum}}\,\underset{\mathrm{helicity}}{\overline{\sum}}}

\title{Full-colour double-virtual amplitudes for associated production of a Higgs boson with a bottom-quark pair at the LHC}

\author[a]{Simon Badger,}
\author[b,c]{Heribertus Bayu Hartanto,}
\author[d]{Rene Poncelet,}
\author[e]{Zihao Wu,}
\author[f,g]{Yang Zhang,}
\author[h]{Simone Zoia}

\affiliation[a]{Dipartimento di Fisica and Arnold-Regge Center, Università di Torino, and INFN, Sezione di Torino, Via P.\ Giuria 1, I-10125 Torino, Italy}
\affiliation[b]{Asia Pacific Center for Theoretical Physics, Pohang, 37673, Korea}
\affiliation[c]{Department of Physics, Pohang University of Science and Technology, Pohang, 37673, Korea}
\affiliation[d]{The Henryk Niewodnicza\'nski Institute of Nuclear Physics, ul.\ Radzikowskiego 152, 31-342 Krakow, Poland}
\affiliation[e]{School of Fundamental Physics and Mathematical Sciences, Hangzhou Institute for Advanced Study, UCAS, Hangzhou, 310000, China}
\affiliation[f]{Interdisciplinary Center for Theoretical Study, University of Science and Technology of China, Hefei, Anhui 230026, China}
\affiliation[g]{Peng Huanwu Center for Fundamental Theory, Hefei, Anhui 230026, China}
\affiliation[h]{Physik-Institut, University of Zurich, Winterthurerstrasse 190, 8057 Zurich, Switzerland}

\emailAdd{simondavid.badger@unito.it}
\emailAdd{bayu.hartanto@apctp.org}
\emailAdd{rene.poncelet@ifj.edu.pl}
\emailAdd{wuzihao@mail.ustc.edu.cn}
\emailAdd{yzhphy@ustc.edu.cn}
\emailAdd{simone.zoia@physik.uzh.ch}

\preprint{
  \begin{tabular}{l}
   IFJPAN-IV-2024-14 \\
   USTC-ICTS/PCFT-24-55 \\
   ZU-TH-63/24
  \end{tabular}
}

\abstract{
We present the double-virtual amplitudes contributing to the production of a
Higgs boson in association with a $b\bar{b}$ pair at the Large Hadron Collider.
We perform the computation within the five-flavour scheme, which employs
massless bottom quarks and finite bottom-Yukawa coupling, taking into account
all the colour structures.  We derive the analytic form of the helicity
amplitudes through finite-field reconstruction techniques.  The analytic
expressions have been implemented in a public \textsc{C++} library, and we
demonstrate that evaluations are sufficiently stable and efficient for use in
phenomenological studies.
}

\begin{document}
\maketitle
\flushbottom

\section{Introduction}

Studying the properties of the Higgs boson is one of the most important tasks in the programme of the 
Large Hadron Collider (LHC) following its discovery in 2012~\cite{ATLAS:2012yve,CMS:2012qbp}.
Precision measurements of the Higgs boson coupling to other particles ---~such as top or bottom quark, $W$ or $Z$ boson~--- will improve our understanding of the Standard Model and might unveil deviations from it~\cite{ATLAS:2022vkf,CMS:2022dwd}, especially in view of the vast amount of data collected from previous and current runs 
as well as those expected from the upcoming high-luminosity phase (HL-LHC)~\cite{Cepeda:2019klc}. 
The production of a Higgs boson in association with a bottom-quark
pair at the LHC ($pp\to b\bar{b}H$) provides a direct access to probe the bottom-quark Yukawa coupling. 

The fully inclusive cross section of $b\bar{b}H$ production at the LHC is comparable to that of Higgs production 
in association with a top-quark pair ($pp\to t\bar{t}H$), which comes after the gluon fusion channel ($pp(gg)\to H$), vector boson fusion ($pp\to qqH$) 
and $W/Z+H$ associated production~\cite{LHCHiggsCrossSectionWorkingGroup:2016ypw}.
Imposing the $b$-jet identification to $b\bar{b}H$ production (by tagging either one or two $b$ jets) significantly decreases the production rate with respect to the inclusive case. In addition, large irreducible backgrounds
may hinder the measurement of bottom-Yukawa coupling through the $b\bar{b}H$ production mechanism~\cite{Pagani:2020rsg}. 
Further studies to extract the $b\bar{b}H$ signal through its kinematic shape, as well as a proposal to probe 
non-standard $b\bar{b}H$ are available in the literature~\cite{Grojean:2020ech,Konar:2021nkk}. 
The strength of the bottom-Yukawa coupling can be significantly amplified in models with modified Higgs sector, such as 
the Two Higgs Doublet Models (2HDM's) and the Minimal Supersymmetric Standard Model (MSSM), 
resulting in the enhancement of the $\bbh$ production cross section~\cite{Balazs:1998nt,Dawson:2005vi}.
In addition to the possibility of constraining beyond-the-Standard-Model (BSM) scenarios that introduce a modification to the bottom-Yukawa coupling, $\bbh$ production also serves
as an irreducible background for double Higgs production, where one of the Higgs boson decays into a bottom-quark pair. The accurate simulation of the $b\bar{b}H$ production 
will thus improve the measurement of the $pp\to HH$ cross section as well as the extraction of the triple Higgs coupling~\cite{Manzoni:2023qaf}. 
The first search for $\bbh$ production at the LHC has been carried out by the CMS collaboration using 13~TeV data with 138$^{-1}$~fb of integrated luminosity, setting an upper limit on the $\bbh$ total 
cross section~\cite{CMS:2024eur}.

Theoretical predictions for the $\bbh$ production can be obtained within either the five-flavour scheme (5FS) or the four-flavour scheme (4FS).
In the 5FS, the bottom quark is treated as massless and can appear in the initial state.
In the 4FS, instead, the bottom quark is treated as massive and is only allowed to appear in the final state.
Owing to the vanishing of the bottom-quark mass, higher-order predictions are easier to obtain in the 5FS than in the 4FS.
Indeed, theoretical predictions for inclusive $\bbh$ production in the 5FS ($pp(b\bar{b})\to H$) are available up to the Next-to-Next-to-Next-to-Leading Order (N3LO) in Quantum Chromodynamics (QCD)~\cite{Dicus:1998hs,Balazs:1998sb,Maltoni:2003pn,Harlander:2003ai,Belyaev:2005bs,Harlander:2010cz,Ozeren:2010qp,Buhler:2012ytl,Harlander:2012pb,Harlander:2014hya,
Ahmed:2014pka,Gehrmann:2014vha,AH:2019xds,Duhr:2019kwi,Mondini:2021nck},
and matching to a parton shower (PS) has been carried out recently at Next-to-Next-to-Leading Order (NNLO) QCD accuracy~\cite{Biello:2024vdh}.
In the case of one-tagged $b$ jet, the 5FS theoretical predictions have been considered by incorporating NLO QCD~\cite{Campbell:2002zm}, weak~\cite{Dawson:2010yz}, 
and supersymmetric QCD~\cite{Dawson:2011pe} corrections.
On the other hand, the theoretical predictions in the 4FS are available up to Next-to-Leading Order (NLO) accuracy 
in QCD~\cite{Dittmaier:2003ej,Dawson:2003kb,Dawson:2004sh}, 
and in supersymmetric QCD~\cite{Liu:2012qu,Dittmaier:2014sva},
including also, in addition to the bottom-quark Yukawa coupling, the contribution from the top-quark Yukawa coupling as well as their interferences~\cite{Deutschmann:2018avk}.
4FS NLO QCD predictions matched to parton shower are also available within the \textsc{MC@NLO} and \textsc{Powheg} matching schemes~\cite{Wiesemann:2014ioa,Jager:2015hka}.
NNLO QCD prediction matched to parton shower for $b\bar{b}H$ production have been recently computed in the 4FS~\cite{Biello:2024pgo} employing approximated two-loop matrix elements derived from the massless amplitudes in the leading colour approximation.

In this work, we compute analytically the two-loop five-particle scattering amplitudes contributing to $pp\to b\bar{b}H$ production in the 5FS, retaining  
the complete colour structure. These amplitudes are required in the NNLO QCD computation of $b\bar{b}H$ production where the two
bottom quarks are identified. In the case of only one tagged bottom quark, instead, they contribute at N3LO QCD accuracy. Finally, for inclusive
$\bbh$ production, they appear in the (Next-to-)$^4$-Leading Order (N4LO) QCD calculation. This work extends the previous result where the two-loop five-particle amplitudes for 
$\bbh$ production were derived in the leading colour approximation~\cite{Badger:2021ega}. 

The computation of two-loop five-point amplitudes has drawn huge interest in the past few years,
since in most cases it is the main bottleneck in obtaining NNLO QCD prediction for $2\to 3$ scattering processes at the LHC. 
Impressive progress was made in this context, owing to several insights.
Firstly, the advancement in the calculation of the required Feynman integrals using the
differential equation method~\cite{Gehrmann:2015bfy,Papadopoulos:2015jft,Abreu:2018rcw,Chicherin:2018mue,Chicherin:2018old,Abreu:2018aqd,Abreu:2020jxa,Canko:2020ylt,Abreu:2021smk,Kardos:2022tpo,Abreu:2023rco} led to the construction of bases of special functions called \textit{pentagon functions}~\cite{Gehrmann:2018yef,Chicherin:2020oor,Chicherin:2021dyp,Abreu:2023rco}, which can be evaluated efficiently and which allow for an efficient computation of the amplitudes.
Secondly, the use of finite-field arithmetic has proven to be critical in tackling the algebraic complexity of multi-particle amplitudes~\cite{vonManteuffel:2014ixa,Peraro:2016wsq,Klappert:2019emp,Peraro:2019svx,Klappert:2020aqs,Klappert:2020nbg}. 
Finally, new techniques to obtain optimised systems of integration-by-part (IBP) relations further reduced one of the main bottlenecks in the computation of scattering amplitudes (see for example Refs.~\cite{Gluza:2010ws,Schabinger:2011dz,
Ita:2015tya,Chen:2015lyz,Bohm:2017qme,Bosma:2018mtf,Boehm:2020zig,Guan:2024byi,Chestnov:2024mnw}). 
Thanks to all these developments, full colour two-loop analytic results~\cite{Badger:2019djh,Agarwal:2021vdh,Badger:2021imn,Abreu:2023bdp,Badger:2023mgf,Agarwal:2023suw,DeLaurentis:2023nss,DeLaurentis:2023izi} 
are now available for all massless five-particle scattering processes of interest for the LHC, superseding the leading colour results~\cite{
Gehrmann:2015bfy,Badger:2018enw,Abreu:2018zmy,Abreu:2019odu,Abreu:2020cwb,Chawdhry:2020for,Agarwal:2021grm,Chawdhry:2021mkw,Abreu:2021oya}.
In the case of two-loop five-particle amplitudes with an external mass, instead, analytic expressions are 
only available in the leading colour approximation~\cite{Badger:2021nhg,Badger:2021ega,Abreu:2021asb,Badger:2022ncb}, although the full set of master integrals have been computed. 
Full colour results for a process of this class have been recently obtained 
for $W\gamma\gamma$ production at the LHC, with the subleading colour
contributions taken into account numerically~\cite{Badger:2024sqv}.
The availability of compact analytic results for two-loop five-point amplitudes, together with a stable and efficient routine for the numerical evaluation of the pentagon functions~\cite{PentagonFunctions:cpp}, have enabled several NNLO QCD calculations for $2\to 3$ scattering processes~\cite{Chawdhry:2019bji,Kallweit:2020gcp,Chawdhry:2021hkp,Czakon:2021mjy,Badger:2021ohm,Chen:2022ktf,
Alvarez:2023fhi,Badger:2023mgf,Hartanto:2022qhh,Hartanto:2022ypo,Buonocore:2022pqq,Catani:2022mfv,Buonocore:2023ljm,Mazzitelli:2024ura,Devoto:2024nhl}.
The result presented in this work is the first analytic full colour amplitude for a two-loop five-point scattering process with an external mass. 

Although efforts to include more external or internal masses, or even more external legs, have recently been 
made~\cite{Henn:2021cyv,Badger:2022hno,FebresCordero:2023pww,Agarwal:2024jyq,Henn:2024ngj,Badger:2024fgb,Jiang:2024eaj,Abreu:2024yit},
computing NNLO QCD predictions for $\bbh$ production in the 4FS with exact two-loop matrix elements remains out of reach of current amplitude technology. However, given that the bottom-quark mass ($m_b$)
is the smallest scale in the process, a good approximation of the 4FS prediction may still be obtained by restoring the leading $m_b$ contributions through the \textit{massification} prescription~\cite{Mitov:2006xs} starting from our massless-$b$ results.
Indeed, in the context of $2 \to 3$ processes, this approach has already
been adopted successfully to $pp\to b\bar{b}W$, $pp\to b\bar{b}Z$ and $pp\to b\bar{b}H$~\cite{Buonocore:2022pqq,Mazzitelli:2024ura,Biello:2024pgo} by employing massless-$b$ leading colour amplitudes.

Our paper is organised as follows. In Section~\ref{sec:bbH_amplitude}, we discuss the full colour structure of the
partonic amplitudes together with the definition of the finite remainders. Section~\ref{sec:amplitude_computation} is devoted to our computational strategy, starting from how we defined the helicity
amplitudes, followed by a discussion of our finite-field framework. 
We present a benchmark numerical evaluation at a physical phase-space point, describe the \textsc{C++} implementation and examine the numerical stability 
of our amplitude in Section~\ref{sec:results}.
We finally draw our conclusion in Section~\ref{sec:conclusion}.

\section{Structure of the amplitudes}
\label{sec:bbH_amplitude}

We compute the two-loop five-particle helicity amplitudes contributing to $\bbh$ production at hadron colliders in the 5FS.
At partonic level, there is a single two-quark two-gluon scattering process
\begin{align}
\label{eq:bbggHdef}
& 0 \rightarrow  \bar{b}(p_1) + b(p_2) + g(p_3) + g(p_4) + H(p_5) \,, 
\end{align}
and two four-quark scattering processes
\begin{align}
\label{eq:bbqqHdef}
& 0 \rightarrow  \bar{b}(p_1) + b(p_2) + \bar{q}(p_3) + q(p_4) + H(p_5) \,,  \\
\label{eq:bbbbHdef}
& 0 \rightarrow  \bar{b}(p_1) + b(p_2) + \bar{b}(p_3) + b(p_4) + H(p_5) \,. 
\end{align}
In this section we first define our parametrisation of the kinematics.
Next, we describe the structure of the two-loop amplitudes for these classes of partonic process.
We then discuss the construction of finite remainders, obtained by subtracting the ultraviolet (UV) and infrared (IR) singularities from the bare amplitudes.
Finally, we give the explicit dependence of the finite remainders on the renormalisation scale.

\subsection{Kinematics}
\label{sec:Kinematics}

We work in the 't Hooft-Veltman (HV) scheme with $d=4-2\eps$ space-time dimensions and four-dimensional external momenta $p_i$.
The latter are taken to be all outgoing, and satisfy momentum conservation,
\begin{equation}
	\sum_{i=1}^{5} p_i = 0 \,,
\end{equation}
as well as the following on-shell conditions:
\begin{align}
	\label{eq:onshellcondition}
	\begin{aligned}
		p_i^2 & = 0 \quad \forall \; i = 1, \dots , 4\,, \\
		p_5^2 & = m_H^2 \,,
	\end{aligned}
\end{align}
where $m_H$ is the Higgs boson mass. Although the bottom-quark mass is set to zero in Eq.~\eqref{eq:onshellcondition}, as required in the 5FS, the bottom-quark Yukawa coupling is kept 
finite~\cite{Harlander:2003ai,Duhr:2019kwi,Mondini:2021nck}. 
We choose the following six independent Mandelstam invariants,
\begin{equation}
	\vec{s} = \lbrace s_{12}, s_{23}, s_{34}, s_{45}, s_{15}, m_H^2 \rbrace \,,
\end{equation}
where $s_{ij \cdots k} = (p_i + p_j + \cdots + p_k)^2$. Together with the following parity-odd invariant,
\begin{equation}
	\tr_5 = 4 \, \i \, \varepsilon_{\mu\nu\rho\sigma} \, p_1^{\mu} p_2^{\nu} p_3^{\rho} p_4^{\sigma} \,,
\end{equation}
where $\varepsilon_{\mu\nu\rho\sigma}$ is the anti-symmetric Levi-Civita pseudo-tensor, they
fully describe the five-particle kinematics in the presence of an external mass.

While the sign of $\tr_5$ captures the parity degree of freedom of the helicity amplitudes, $\tr_5^2 \equiv \Delta_5$ is a degree-4 irreducible polynomial in the Mandelstam invariants $\vec{s}$. The parametrisation of the kinematics in terms of $\vec{s}$ therefore contains the square root of $\Delta_5$.
For the computation of the helicity amplitudes, we prefer to adopt a parametrisation of the kinematics in terms of momentum-twistor variables~\cite{Hodges:2009hk,Badger:2016uuq}, which rationalises $\sqrt{\Delta_5}$ and the spinor-helicity brackets $\spA{i}{j}$ and $\spB{i}{j}$ using a minimal set of variables (see e.g.\ Ref.~\cite{Badger:2023eqz} for an introduction).

In particular, we employ the parametrisation of Ref.~\cite{Badger:2021ega}.
The Mandelstam invariants $\vec{s}$ are expressed in terms of the momentum-twistor variables $\vec{x} = \{ x_1, \ldots, x_6 \}$ as
\begin{align}
\label{eq:sijtox}
\begin{aligned}
s_{12} & = x_1 \,, \\
s_{23} & = x_1 x_4\,, \\
s_{34} & = x_1 \left[ \frac{(1+x_3) x_4}{x_2} - x_3 \Bigl(1 + x_4 (x_5 - 1) \Bigr)\right]\,, \\
s_{45} & = x_1 x_6\,, \\
s_{15} & = x_1 x_3 (x_2 - x_4 x_5)\,, \\
m_H^2 & = \frac{x_1 x_3}{x_4} (x_2 - x_4) \Bigl[ x_4 (x_5 - 1) + x_6 \Bigr] \,.
\end{aligned}
\end{align}
Conversely, the expression of the momentum-twistor variables in terms of momenta requires also the parity degree of freedom, which is captured by Dirac traces with $\gamma_5$:
\begin{equation}
\begin{alignedat}{2}
  & x_1 = s_{12} \,, \qquad && x_2 = -\frac{\trp(1234)}{s_{12}s_{34}} \,, \\
  & x_3 = \frac{\trp[134152]}{s_{13}\trp{[1452]}} \,, \qquad && x_4 = \frac{s_{23}}{s_{12}} \,, \\
  & x_5 = -\frac{\trm[1(2+3)(1+5)523]}{s_{23}\trm[1523]} \,, \qquad && x_6 = \frac{s_{123}}{s_{12}} \,,
\end{alignedat}
\label{eq:momtwistor5pt}
\end{equation}
where
\begin{align}
& \tr_{\pm}(ij \cdots kl) = \frac{1}{2} \tr\left[(1\pm\gamma_5)\slashed{p}_i\slashed{p}_j \cdots\slashed{p}_k\slashed{p}_l\right] \,, \\
& \tr_{\pm}[\cdots (i+j) \cdots] = \tr_{\pm}(\cdots i \cdots) + \tr_{\pm}(\cdots j \cdots) \,.
\end{align}

\subsection{Two-quark two-gluon scattering channel}

The colour decomposition of the $L$-loop amplitude $\cM^{(L)}$ for the $\bbggH$ channel is given by
\begin{align}
\label{eq:bbggHcolourdecomposition}
\begin{aligned}
\cM^{(L)}(1_{\bar{b}},2_b,3_g,4_g,5_H) = \; & n^L \bar{g}_s^2 \bar{y}_b \; \bigg\lbrace
(t^{a_3}t^{a_4})_{i_2}^{\;\;\bar{i}_1} \cA^{(L)}_{34}(1_{\bar{b}},2_b,3_g,4_g,5_H)  \\
& + (t^{a_4}t^{a_3})_{i_2}^{\;\;\bar{i}_1} \cA^{(L)}_{43}(1_{\bar{b}},2_b,3_g,4_g,5_H)
  + \delta_{i_2}^{\;\;\bar{i}_1} \delta^{a_3 a_4} \cA^{(L)}_{\delta}(1_{\bar{b}},2_b,3_g,4_g,5_H)
\bigg\rbrace \,,
\end{aligned}
\end{align}
where $n = (4\pi)^\eps e^{-\eps \gamma_E}\,\bar\alpha_s/(4\pi)$, $\bar\alpha_s=\bar{g}_s^2/(4\pi)$, $\bar{g}_s$ is the bare strong coupling constant,
$\bar{y}_b$ is the bare bottom-Yukawa coupling,
and $t^a$ are the $SU(N_c)$ fundamental generators satisfying $\tr(t^{a}t^{b}) = \delta^{ab}/2$.
The $\cA^{(L)}_{43}$ partial amplitudes can be obtained from the $\cA^{(L)}_{34}$ ones through
\begin{equation}
\cA^{(L)}_{43}(1_{\bar{b}},2_b,3_g,4_g,5_H) = \cA^{(L)}_{34}(1_{\bar{b}},2_b,4_g,3_g,5_H) \,.
\end{equation}
We thus compute only  $\cA^{(L)}_{34}$ and $\cA^{(L)}_{\delta}$.
We further note that $\cA^{(L)}_{\delta}$ vanishes at tree level.

\begin{figure}[t!]
  \begin{center}
    \includegraphics[width=0.985\textwidth]{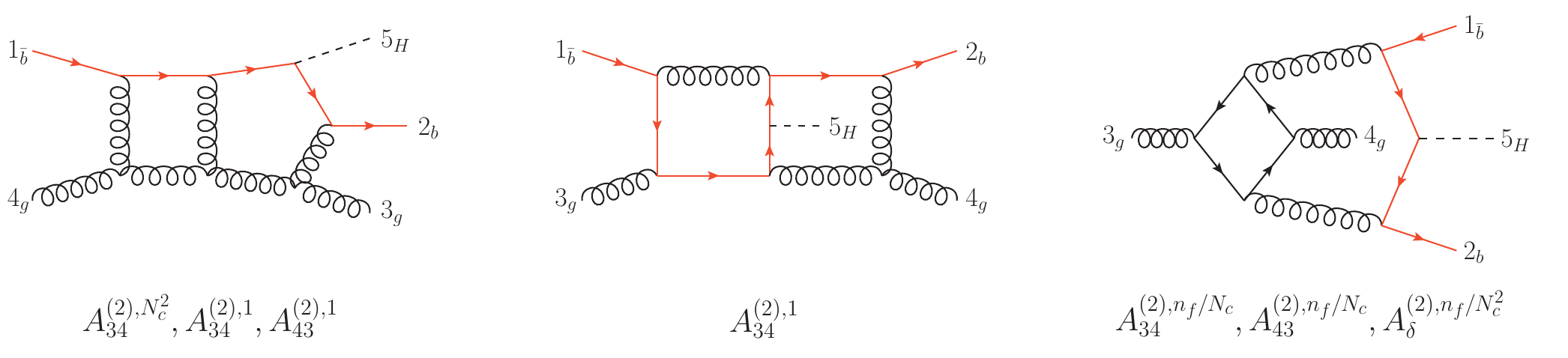}
  \end{center}
    \caption{Representative two-loop Feynman diagrams appearing in the $\bbggH$ scattering process, together with the partial amplitudes associated with each of the diagrams. Dashed lines denote the Higgs boson, while solid lines denote massless quarks (red for the bottom quark).}
  \label{fig:bbggH2L}
\end{figure}

The partial amplitudes are decomposed in terms of the number of colours ($N_c$) and the number of light quark flavours ($n_f$) appearing in closed massless quark loops.
The $(N_c,n_f)$ decomposition of the one-loop partial amplitudes is given~by
\begin{subequations}
\label{eq:bbggHNcNf1Ldecomposition}
  \begin{align}
    \cA^{(1)}_{34} & = N_c A^{(1),N_c}_{34} + \frac{1}{N_c} A^{(1),1/N_c}_{34} + n_f  A^{(1),n_f}_{34}\,, \\
    \cA^{(1)}_{\delta} &= A^{(1),1}_{\delta} \,,
  \end{align}
\end{subequations}
while at two loops we have
\begin{subequations}
\label{eq:bbggHNcNf2Ldecomposition}
  \begin{align}
    \cA^{(2)}_{34} & =  N_c^2 A^{(2),N_c^2}_{34} + A^{(2),1}_{34} + \frac{1}{N_c^2} A^{(2),1/N_c^2}_{34}
                      + N_c n_f  A^{(2),N_c n_f}_{34}  + \frac{n_f}{N_c}  A^{(2),n_f/N_c}_{34} + n_f^2  A^{(2),n_f^2}_{34 }\,, \\
    \cA^{(2)}_{\delta} & =   N_c A^{(2),N_c}_{\delta} + \frac{1}{N_c} A^{(2),1/N_c}_{\delta}
                           + n_f A^{(2),n_f}_{\delta} + \frac{n_f}{N_c^2} A^{(2),n_f/N_c^2}_{\delta} \,.
  \end{align}
\end{subequations}
Representative two-loop Feynman diagrams together with the partial amplitudes they contribute to are shown in Fig.~\ref{fig:bbggH2L}.

There are 8 distinct helicity configurations, of which 3 are independent and have been chosen~as
\begin{align}
\label{eq:indephelgg}
\begin{aligned}
\cM^{(L)}(1_{\bar{b}}^+,2_b^+,3_g^+,4_g^+,5_H) \,,  \\
\cM^{(L)}(1_{\bar{b}}^+,2_b^+,3_g^+,4_g^-,5_H) \,,  \\
\cM^{(L)}(1_{\bar{b}}^+,2_b^+,3_g^-,4_g^-,5_H) \,.
\end{aligned}
\end{align}

We obtain the full set of partial amplitudes and helicity configurations from the independent ones by means of permutations of the external legs as well as parity conjugation.
Since the momentum-twistor parametrisation of the kinematics fixes all spinor phases in order to minimise the number of independent variables, the phase information has to be restored before permutations or parity conjugation can be carried out.
For this purpose, we employ the following phase factors:
\begin{subequations}
\label{eq:bbggHphase}
\begin{align}
	\Phi_{\bbggh}^{++++} & = \frac{s_{12}}{\spA{2}{3}\spA{3}{4}\spA{4}{1}} \,, \\
	\Phi_{\bbggh}^{+++-} & = \frac{\spB{1}{3}^2}{\spB{1}{4}\spB{3}{4}\spA{2}{3}} \,, \\
	\Phi_{\bbggh}^{++--} & = \frac{\spB{1}{2}^2}{\spB{2}{3}\spB{3}{4}\spB{4}{1}} \,.
\end{align}
\end{subequations}
We refer to Appendix~C of Ref.~\cite{Badger:2023mgf} for a thorough discussion of how to permute and conjugate expressions in terms of momentum-twistor variables.

\subsection{Four-quark scattering channel}

\begin{figure}[t!]
  \begin{center}
    \includegraphics[width=0.985\textwidth]{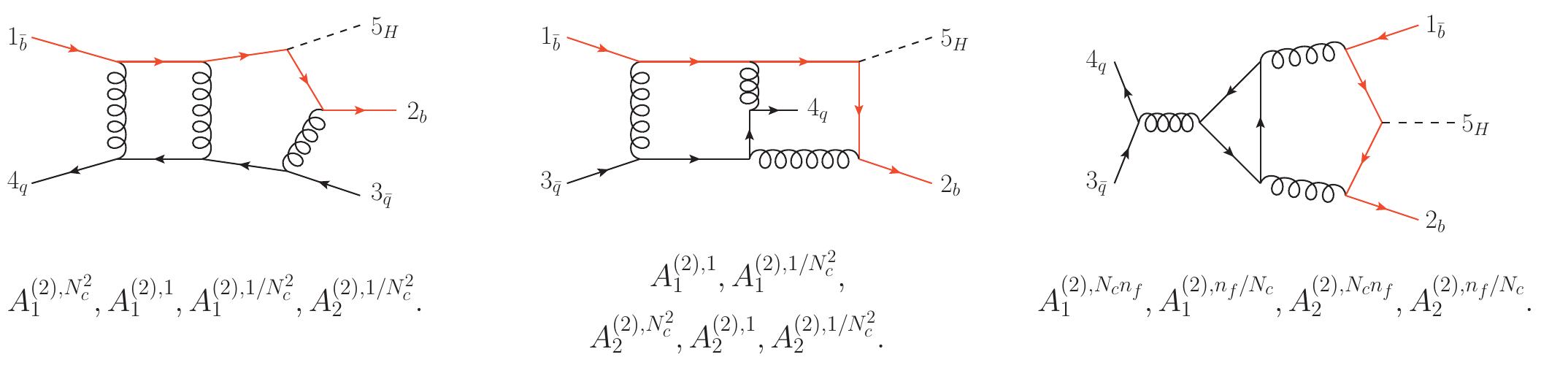}
  \end{center}
  \caption{Representative two-loop Feynman diagrams appearing in the $\bbqqH$ scattering process, together with the partial amplitudes they contribute to. The notation is the same as in Fig.~\ref{fig:bbggH2L}.}
  \label{fig:bbqqH2L}
\end{figure}

For this class of partonic processes we first focus on the case involving two distinct quark flavours: a bottom quark and a massless quark $q \in \{ u,d,c,s\}$.
The colour decomposition of the $L$-loop amplitude $\cM^{(L)}$ for the $\bbqqH$ channel is given by
\begin{align}
\label{eq:bbqqHcolourdecomposition}
\begin{aligned}
\cM^{(L)}(1_{\bar{b}},2_b,3_{\bar{q}},4_q,5_H)  = \; n^L \bar{g}_s^2 \bar{y}_b \; \bigg\lbrace &
    \delta_{i_4}^{\;\;\bar{i}_1} \delta_{i_2}^{\;\;\bar{i}_3} \, \cA^{(L)}_{1}(1_{\bar{b}},2_b,3_{\bar{q}},4_q,5_H)  \\
& + \frac{1}{N_c} \delta_{i_2}^{\;\;\bar{i}_1} \delta_{i_4}^{\;\;\bar{i}_3} \, \cA^{(L)}_{2}(1_{\bar{b}},2_b,3_{\bar{q}},4_q,5_H)
\bigg\rbrace \,.
\end{aligned}
\end{align}
Only one partial amplitude is independent at the tree level, since
\begin{equation}
	\cA^{(0)}_{2}(1_{\bar{b}},2_b,3_{\bar{q}},4_q,5_H) = - \cA^{(0)}_{1}(1_{\bar{b}},2_b,3_{\bar{q}},4_q,5_H) \,. \nonumber
\end{equation}
At higher loop order we instead have to compute both $\cA^{(L)}_{1}$ and $\cA^{(L)}_{2}$.
Their $(N_c,n_f)$ decomposition at one and two loops is given by
\begin{subequations}
\label{eq:bbqqHNcNfdecomposition}
  \begin{align}
	  \cA^{(1)}_{\iamp} & = N_c A^{(1),N_c}_{\iamp} + \frac{1}{N_c} A^{(1),1/N_c}_{\iamp} + n_f  A^{(1),n_f}_{\iamp}\,, \\
	  \cA^{(2)}_{\iamp} & =  N_c^2 A^{(2),N_c^2}_{\iamp} + A^{(2),1}_{\iamp} + \frac{1}{N_c^2} A^{(2),1/N_c^2}_{\iamp}
                      		+ N_c n_f  A^{(2),N_c n_f}_{\iamp}  + \frac{n_f}{N_c}  A^{(2),n_f/N_c}_{\iamp} + n_f^2  A^{(2),n_f^2}_{\iamp}\,,
  \end{align}
\end{subequations}
where $\iamp \in \{ 1,2 \}$. Sample two-loop Feynman diagrams for the $\bbqqH$ 
scattering process are depicted in Fig.~\ref{fig:bbqqH2L} together with the partial amplitudes they contribute to.
From the 4 contributing helicity configurations, only one is independent and was chosen to be
\begin{equation}
\label{eq:indephelqq}
\cM^{(L)}(1_{\bar{b}}^+,2_b^+,3_{\bar{q}}^+,4_q^-,5_H).
\end{equation}
The corresponding spinor phase factor is
\begin{align}
\label{eq:bbqqHphase}
\Phi_{\bbqqh}^{+++-} & = \frac{\spA{4}{2}}{\spA{1}{2}\spA{2}{3}} \,.
\end{align}

We now extend the discussion to the four-quark scattering process involving two bottom-quark pairs: $\bbbbH$.
In this case, the $L$-loop colour decomposition is given by
\begin{align}
\label{eq:bbbbHcolourdecomposition}
\begin{aligned}
\cM^{(L)}(1_{\bar{b}},2_b,3_{\bar{b}},4_b,5_H)  = \; n^L \bar{g}_s^2 \bar{y}_b \; \bigg\lbrace &
    \delta_{i_4}^{\;\;\bar{i}_1} \delta_{i_2}^{\;\;\bar{i}_3} \cA^{(L)}_{1}(1_{\bar{b}},2_b,3_{\bar{q}},4_q,5_H) \\
& + \frac{1}{N_c} \delta_{i_2}^{\;\;\bar{i}_1} \delta_{i_4}^{\;\;\bar{i}_3} \cA^{(L)}_{2}(1_{\bar{b}},2_b,3_{\bar{q}},4_q,5_H)  \\
& + \delta_{i_2}^{\;\;\bar{i}_1} \delta_{i_4}^{\;\;\bar{i}_3} \cA^{(L)}_{3}(1_{\bar{b}},2_b,3_{\bar{q}},4_q,5_H)   \\
& + \frac{1}{N_c} \delta_{i_4}^{\;\;\bar{i}_1} \delta_{i_2}^{\;\;\bar{i}_3} \cA^{(L)}_{4}(1_{\bar{b}},2_b,3_{\bar{q}},4_q,5_H)
\bigg\rbrace \,,
\end{aligned}
\end{align}
where the partial amplitudes $\cA^{(L)}_{3}$ and $\cA^{(L)}_{4}$ can be related to the ones appearing already in the $\bbqqH$ process through the following relations:
\begin{subequations}
\begin{align}
\cA^{(L)}_{3}(1_{\bar{b}},2_b,3_{\bar{q}},4_q,5_H) & = -\cA^{(L)}_{1}(1_{\bar{b}},4_q,3_{\bar{q}},2_b,5_H) \,, \\
\cA^{(L)}_{4}(1_{\bar{b}},2_b,3_{\bar{q}},4_q,5_H) & = -\cA^{(L)}_{2}(1_{\bar{b}},4_q,3_{\bar{q}},2_b,5_H) \,.
\end{align}
\end{subequations}

\subsection{Renormalisation and infrared subtraction}
\label{sec:UVIRsubtraction}

The construction of the finite remainders requires the removal of UV and IR divergences. To obtain the UV renormalisation counterterms, the bare strong coupling
constant $\bar{\alpha}_s$ and bottom-Yukawa coupling $\bar{y}_b$ are replaced by their renormalised counterparts $\alpha_s$ and $y_b$ through the following relations~\cite{Mondini:2019vub},
\begin{align}
	\bar{\alpha}_s \mu_0^{2\eps} S_\eps & =  \alpha_s \mu^{2\eps} \, \bigg\lbrace 1 - \frac{\alpha_s}{4\pi} \frac{\beta_0}{\eps}
					     + \left(\frac{\alpha_s}{4\pi}\right)^2 \left( \frac{\beta_0^2}{\eps^2} - \frac{\beta_1}{2\eps}\right) + \cO(\alpha_s^3) \bigg\rbrace \,, \\
	\bar{y}_b & = y_b \, \bigg\lbrace 1 - \frac{\alpha_s}{4\pi} \frac{3C_F}{\eps} 
		  +  \left(\frac{\alpha_s}{4\pi}\right)^2 \left[ \frac{3}{8\eps^2} \left( 3 C_F^2 + \beta_0 C_F \right) \right. \nonumber \\ 
		  & \qquad\quad \left. - \frac{1}{8\eps} \left( \frac{3}{2} C_F^2 + \frac{97}{6} C_F C_A - \frac{10}{3} C_F T_R n_f \right) \right] + \cO(\alpha_s^3) \bigg\rbrace \,,
\end{align}
where $S_{\eps} = e^{-\eps\gamma_E}/(4\pi)^\eps$ and $\mu$ is the renormalisation scale. The first two $\beta$-function coefficients are
\begin{subequations}
\begin{align}
	\beta_0 & = \frac{11}{3} C_A - \frac{4}{3} T_R n_f \,, \\
	\beta_1 & = \frac{34}{3} C_A^2 - \frac{20}{3} C_A T_R n_f - 4 C_F T_R n_f \,, 
\end{align}
\end{subequations}
with $C_F = (N_c^2-1)/(2N_c)$, $C_A=N_c$ and $T_R=1/2$. The IR singularities of the one- and two-loop amplitudes are known 
universally~\cite{Catani:1998bh,Becher:2009qa,Becher:2009cu,Gardi:2009qi}, and in this work
we adopt the $\overline{\mathrm{MS}}$ scheme according to Refs.~\cite{Becher:2009qa,Becher:2009cu}. 
We note that the present choice of IR subtraction scheme is different from that of the previous leading colour computation, where 
the Catani scheme~\cite{Catani:1998bh} was instead used.
While the tree-level colour dressed finite remainder $\cR^{(0)}$ is simply the bare tree-level amplitude, 
the one- and two-loop colour-dressed finite remainders $\cR^{(1)}$ and $\cR^{(2)}$ are instead obtained by 
subtracting the UV and IR divergences from the bare amplitudes as 
\begin{subequations}
\label{eq:finiteremainder}
\begin{align}
	\big|\cR^{(0)}_\iamp\big> & =  \big|\cM^{(0)}_\iamp\big>  \,, \\
	\big|\cR^{(1)}_\iamp\big> & =  \big|\cM^{(1)}_\iamp\big> - \left( \frac{\beta_0}{\eps} + \frac{3C_F}{\eps} \right) \big|\cM^{(0)}_\iamp\big>  
	                             - \textbf{Z}^{(1)}_\iamp \big|\cM^{(0)}_\iamp\big>  \,, \\
	\big|\cR^{(2)}_\iamp\big> & =  \big|\cM^{(2)}_\iamp\big> - \left( \frac{2\beta_0}{\eps} + \frac{3 C_F}{\eps}\right)  \big|\cM^{(1)}_\iamp\big> 
                                  + \left\lbrace \frac{\beta_0^2}{\eps^2} - \frac{\beta_1}{2\eps} + \frac{3}{8\eps^2} \left( 3 C_F^2 + \beta_0 C_F \right) \right. \nonumber \\
				& \qquad  \left.  - \frac{1}{8\eps} \bigg( \frac{3}{2} C_F^2 + \frac{97}{6} C_F C_A - \frac{10}{3} C_F T_R n_f \bigg) \right\rbrace \big|\cM^{(0)}_\iamp\big> 
				  - \textbf{Z}^{(2)}_\iamp \big|\cM^{(0)}_\iamp\big> - \textbf{Z}^{(1)}_\iamp \big|\cR^{(1)}_\iamp\big>  \,, 
\end{align}
\end{subequations}
where $\iamp \in \{\gg,\qq\}$,
with $\gg$ and $\qq$ shorthands for $\bbggH$ and $\bbqqH$, respectively.
$\big| \cM^{(L)}_{\gg/\qq} \big>$ is a vector in colour space given by
\begin{subequations}
\begin{align}
\big| \cM^{(L)}_\gg \big> & = 
	\begin{pmatrix}  
		\cA^{(L)}_{34}(1_{\bar{b}},2_b,3_g,4_g,5_H) \\ 
		\cA^{(L)}_{43}(1_{\bar{b}},2_b,3_g,4_g,5_H) \\ 
		\cA^{(L)}_{\delta}(1_{\bar{b}},2_b,3_g,4_g,5_H) 
	\end{pmatrix} \,, \\
\big| \cM^{(L)}_\qq \big> & = 
	\begin{pmatrix}  
		\cA^{(L)}_{1}(1_{\bar{b}},2_b,3_{\bar{q}},4_q,5_H) \\ 
		\frac{1}{N_c}\cA^{(L)}_{2}(1_{\bar{b}},2_b,3_{\bar{q}},4_q,5_H) 
	\end{pmatrix} \,,
\end{align}
\end{subequations}
and similarly for $\big|\cR^{(L)}_{\gg/\qq}\big>$.
The IR singularities of the one- and two-loop amplitudes are encoded in the following process-dependent operator in colour space,
\begin{equation}
\label{eq:Zexpansion}
	\textbf{Z}_{\iamp} = \textbf{1} + \frac{\alpha_s}{4\pi} \textbf{Z}^{(1)}_\iamp 
  + \left(\frac{\alpha_s}{4\pi}\right)^2 \textbf{Z}^{(2)}_\iamp + \cO\left(\alpha_s^3\right),\qquad\qquad \iamp \in \{ \gg,\qq \} \,,
\end{equation}
where $\textbf{1}$ is the identity operator. We refer to Refs.~\cite{Becher:2009qa,Becher:2009cu} 
for the definition of $\textbf{Z}_\iamp$. We provide the explicit expressions of $\textbf{Z}^{(1)}_\iamp$ and $\textbf{Z}^{(2)}_\iamp$ for both the $\bbggH$ and $\bbqqH$ processes in our ancillary files. Note that we identify the IR factorisation scale appearing in $\textbf{Z}_\iamp$ with the renormalisation scale $\mu$.

The decomposition of the colour-dressed finite remainders $\cR^{(L)}$ into partial finite remainders $\cF^{(L)}$, 
followed by further decomposition into the $(N_c,n_f)$ components $F^{(L),i}$, follow those of the bare amplitudes 
by replacing $\cM \rightarrow \cR$, $\cA \rightarrow \cF$ and $A \rightarrow F$ in 
Eqs.~\eqref{eq:bbggHcolourdecomposition},~\eqref{eq:bbggHNcNf1Ldecomposition},~\eqref{eq:bbggHNcNf2Ldecomposition},~\eqref{eq:bbqqHcolourdecomposition}~and~~\eqref{eq:bbqqHNcNfdecomposition},
with the same superscripts and subscripts.
In addition, for both the $\bbggH$ and $\bbqqH$ finite remainder colour decompositions in Eqs.~\eqref{eq:bbggHcolourdecomposition}~and~\eqref{eq:bbqqHcolourdecomposition}, 
renormalised strong coupling constant ($g_s$) and bottom-Yukawa coupling ($y_b$) are used in place of the unrenormalised ones.

Finally, we define the tree-level, one-loop and two-loop hard functions as 
\begin{subequations}
\label{eq:hardfunctions}%
\begin{align}
\cH^{(0)} & = \colhelsum \big|\cR^{(0)}\big|^2\,, \\
\cH^{(1)} & = 2 \, \mathrm{Re} \, \colhelsum \cR^{(0)*} \cR^{(1)}\,, \\
\cH^{(2)} & = 2 \, \mathrm{Re} \,  \colhelsum \cR^{(0)*} \cR^{(2)}
             + \colhelsum \big|\cR^{(1)} \big|^2 \,,
\end{align}
\end{subequations}
where the overline denotes the average over the initial state colour and helicity. 
The one-loop finite remainders entering the one- and two-loop hard functions are required to be evaluated only up to order $\eps^0$.
The colour summed $L_1$-loop finite remainder interfered with the $L_2$-loop finite remainder is given for $\bbggH$ by
\begin{align}
\label{eq:bbggHsquaredamplitude}
\begin{aligned}
& \sum_{\mathrm{colour}} \left[\cR^{(L_1)}(1_{\bar{b}},2_b,3_g,4_g,5_H)\right]^* \cR^{(L_2)}(1_{\bar{b}},2_b,3_g,4_g,5_H) \\
& \qquad\qquad = \; g_s^4 y_b^2 n^{L_1} n^{L_2} (N_c^2-1)
  \bigg\lbrace  \frac{N_c^2-1}{4 N_c} \left[ \cF^{(L_1)*}_{34} \cF^{(L_2)}_{34} + \cF^{(L_1)*}_{43} \cF^{(L_2)}_{43} \right]  \\
&\qquad\qquad\qquad  -\frac{1}{4 N_c} \left[ \cF^{(L_1)*}_{34} \cF^{(L_2)}_{43} + \cF^{(L_1)*}_{43} \cF^{(L_2)}_{34} \right]
  + N_c  \cF^{(L_1)*}_{\delta} \cF^{(L_2)}_{\delta} \\
&\qquad\qquad\qquad  +\frac{1}{2} \left[
  \cF^{(L_1)*}_{34} \cF^{(L_2)}_{\delta}
+ \cF^{(L_1)*}_{43} \cF^{(L_2)}_{\delta}
+ \cF^{(L_1)*}_{\delta} \cF^{(L_2)}_{34}
+ \cF^{(L_1)*}_{\delta} \cF^{(L_2)}_{43}
\right] \bigg\rbrace\,,
\end{aligned}
\end{align}
while for $\bbqqH$ it is given by
\begin{align}
\label{eq:bbqqHsquaredamplitude}
\begin{aligned}
& \sum_{\mathrm{colour}} \left[\cR^{(L_1)}(1_{\bar{b}},2_b,3_{\bar{q}},4_q,5_H)\right]^* \cR^{(L_2)}(1_{\bar{b}},2_b,3_{\bar{q}},4_q,5_H) \\
& \qquad\qquad = \; g_s^4 y_b^2 n^{L_1} n^{L_2}  \bigg\lbrace
  N_c^2  \cF^{(L_1)*}_{1} \cF^{(L_2)}_{1}
+   \cF^{(L_1)*}_{2} \cF^{(L_2)}_{2}
+   \cF^{(L_1)*}_{1} \cF^{(L_2)}_{2}
+   \cF^{(L_1)*}_{2} \cF^{(L_2)}_{1}  \bigg\rbrace\,.
\end{aligned}
\end{align}
The colour summed $L_1$-loop finite remainder interfered with the $L_2$-loop finite remainder for $\bbbbH$ can be obtained from that of $\bbqqH$
in Eq.~\eqref{eq:bbqqHsquaredamplitude} by applying the following replacements:
\begin{equation}
\cF^{(L)}_{1} \rightarrow \cF^{(L)}_{1} + \frac{1}{N_c} \cF^{(L)}_{4} \qquad \mathrm{and} \qquad
\cF^{(L)}_{2} \rightarrow N_c \left( \cF^{(L)}_{3} + \frac{1}{N_c} \cF^{(L)}_{2}\right) \,.
\end{equation}

\subsection{$\mu$ dependence of the finite remainder}
\label{sec:mudependence}

We compute the finite remainders $\cR^{(L)}_{\gg/\qq}$ with the renormalisation scale $\mu$ set to $1$ for the sake of simplicity.
The full $\mu$ dependence is obtained from the finite remainders with $\mu=1$ through `$\mu$-restoring terms' $\delta\mathcal{R}^{(L)}_\iamp(\vec{x}, \mu^2)$,
\begin{align}
| \mathcal{R}^{(L)}_\iamp(\vec{x} , \mu^2)\rangle = | \mathcal{R}^{(L)}_\iamp(\vec{x} , \mu^2=1)\rangle + | \delta\mathcal{R}^{(L)}_\iamp(\vec{x}, \mu^2)\rangle\,, 
	\qquad \iamp \in \{\gg,\qq\}  \,.
\end{align}
Since $\mu$ enters the finite remainders through the renormalised couplings ($\alpha_s(\mu)$ and $y_b(\mu)$) and the IR-pole operator $\textbf{Z}_{\gg/\qq}$, the $\mu$-restoring terms depend only on UV/IR-subtraction ingredients and lower-order finite remainders with $\mu=1$.
The one-loop $\mu$-restoring term is given by
\begin{align} 
\label{eq:mu1L}
\begin{aligned}
& | \delta\mathcal{R}^{(1)}_\iamp(\vec{x},\mu^2) \rangle =  \left[ \frac{1}{2} \log^2(\mu^2) \, \textbf{Z}^{(1)}_{\iamp;-2} +
  \log(\mu^2) \left(\beta_0 \, \textbf{1}  + 3 C_F \, \textbf{1}   +\textbf{Z}^{(1)}_{\iamp;-1}(\vec{s}) \right) \right] | \mathcal{M}^{(0)}_\iamp(\vec{x}) \rangle \,, \\
\end{aligned}
\end{align}
while the two-loop $\mu$-restoring term is
\begin{align} 
\label{eq:mu2L} 
\begin{aligned}
& | \delta\mathcal{R}^{(2)}_\iamp(\vec{x},\mu^2) \rangle = \left[ 
	\frac{1}{2} \log^2(\mu^2)  \, \textbf{Z}^{(1)}_{\iamp;-2} + \log(\mu^2) \left(2 \beta_0 \,  \textbf{1} + 3 C_F \,  \textbf{1}  
	+ \textbf{Z}^{(1)}_{\iamp;-1}(\vec{s}) \right) \right]  
	| \mathcal{R}^{(1)}_\iamp(\vec{x},\mu^2=1) \rangle  \\
 & \phantom{ |  \delta\mathcal{R}^{(2)}_\iamp(\vec{x},\mu^2) \rangle = } + 
\biggl\{ \log^4(\mu^2) \left[-\frac{5}{24}  \textbf{Z}^{(1)}_{\iamp;-2} \cdot \textbf{Z}^{(1)}_{\iamp;-2} +\frac{2}{3} \, \textbf{Z}^{(2)}_{\iamp;-4} \right]  \\ 
 & \phantom{ | \delta\mathcal{R}^{(2)}(\vec{s},\mu^2) \rangle = + \biggl\{} +
   \frac{1}{6} \log^3(\mu^2) \left[10 \beta_0 \, \textbf{Z}^{(1)}_{\iamp;-2} + 9 C_F \, \textbf{Z}^{(1)}_{\iamp;-2} 
	- 5 \, \textbf{Z}^{(1)}_{\iamp;-2} \cdot \textbf{Z}^{(1)}_{\iamp;-1}(\vec{s}) + 8 \, \textbf{Z}^{(2)}_{\iamp;-3}(\vec{s}) \right]  \\
 & \phantom{ |  \delta\mathcal{R}^{(2)}(\vec{s},\mu^2) \rangle = + \biggl\{} +
	\log^2(\mu^2) \left[\beta_0^2 \, \textbf{1} + \frac{9}{2} C_F^2 \, \textbf{1} + \frac{9}{2} C_F \beta_0 \, \textbf{1} + 3 C_F \, \textbf{Z}^{(1)}_{\iamp;-1}(\vec{s})
	+\frac{5}{2} \beta_0 \, \textbf{Z}^{(1)}_{\iamp;-1}(\vec{s}) \right.  \\
 & \phantom{ |  \delta\mathcal{R}^{(2)}(\vec{s},\mu^2) \rangle = + \biggl\{} 
	\qquad\qquad\qquad \left.
	-\frac{1}{2} \textbf{Z}^{(1)}_{\iamp;-1}(\vec{s}) \cdot \textbf{Z}^{(1)}_{\iamp;-1}(\vec{s}) +2 \, \textbf{Z}^{(2)}_{\iamp;-2}(\vec{s}) \right]  \\
	& \phantom{ | \delta\mathcal{R}^{(2)}_\iamp(\vec{s},\mu^2) \rangle = + \biggl\{} + \log(\mu^2) \left[ \beta_1 \, \textbf{1} +2 \, \textbf{Z}^{(2)}_{\iamp;-1}(\vec{s}) \right] \biggr\}  | \mathcal{M}^{(0)}_\iamp(\vec{x}) \rangle \,,
\end{aligned}
\end{align}
for $\iamp \in \{\gg,\qq\}$.
Here, $\textbf{Z}^{(L)}_{\iamp;k}$ denotes the order-$\eps^k$ term in the Laurent expansion of the pole operator $\textbf{Z}^{(L)}_{\iamp}$ around $\eps=0$ with $\mu=1$, and we give the arguments of all functions explicitly to highlight that all logarithms of $\mu$ are accounted for.
The leading poles of $\textbf{Z}^{(L)}_{\iamp}$, i.e.\ $\textbf{Z}^{(1)}_{\iamp;-2}$ and $\textbf{Z}^{(2)}_{\iamp;-4}$, are constant, while the higher-order terms contain logarithms of $-s_{ij}$ which are analytically continued to the relevant phase-space region.

\section{Amplitude calculation}
\label{sec:amplitude_computation}

In this section we discuss our strategy to obtain the analytic expression of the helicity amplitudes for $pp\to b\bar{b}H$ up to two-loop order. We first describe
the construction of the helicity amplitudes by using the projectors method, followed by a discussion of our computational toolchain based on finite-field
arithmetic within the \textsc{FiniteFlow} framework~\cite{Peraro:2019svx}.
We conclude with a number of checks that we have performed to validate our results.

\subsection{Tensor decomposition and helicity amplitudes}
\label{sec:tensordecomposition}

We begin the construction of the helicity amplitudes by decomposing the $\bbggH$ and $\bbqqH$ amplitudes into sets of independent tensor structures
$T_{\gg/\qq,i}$~\cite{Peraro:2019cjj,Peraro:2020sfm}.
Such tensor decompositions are valid at any loop order and for any partial sub-amplitudes.
To improve the readability, we suppress the indices associated with the colour factor and the ($N_c$,$n_f$) contributions.
In what follows therefore $A^{(L)}_{\gg/\qq}$ stands for any term in the decomposition of the partial amplitudes
appearing in Eqs.~\eqref{eq:bbggHNcNf1Ldecomposition},~\eqref{eq:bbggHNcNf2Ldecomposition}~and~\eqref{eq:bbqqHNcNfdecomposition}.
The tensor decompositions are then given by
\begin{subequations}
\begin{align}
	A^{(L)}_{\gg} & = \sum_{i=1}^{8} T_{\gg,i} \, \omega^{(L)}_{\gg,i} \,,  \label{eq:bbggHtensordecomposition}\\
	A^{(L)}_{\qq} & = \sum_{i=1}^{4} T_{\qq,i} \, \omega^{(L)}_{\qq,i} \,,  \label{eq:bbqqHtensordecomposition}
\end{align}
\end{subequations}
where
\begin{align}
\label{eq:bbggHtensorstructure}
\begin{aligned}
T_{\gg,1} & = \bar{u}(p_2) v(p_1) \; p_1 \cdot \vareps(p_3,q_3) \; p_1 \cdot \vareps(p_4,q_4) \,,  \\
T_{\gg,2} & = \bar{u}(p_2) v(p_1) \; p_1 \cdot \vareps(p_3,q_3) \; p_2 \cdot \vareps(p_4,q_4) \,,  \\
T_{\gg,3} & = \bar{u}(p_2) v(p_1) \; p_2 \cdot \vareps(p_3,q_3) \; p_1 \cdot \vareps(p_4,q_4) \,,  \\
T_{\gg,4} & = \bar{u}(p_2) v(p_1) \; p_2 \cdot \vareps(p_3,q_3) \; p_2 \cdot \vareps(p_4,q_4) \,,  \\
T_{\gg,5} & = \bar{u}(p_2) \slashed{p}_3\slashed{p}_4 v(p_1) \; p_1 \cdot \vareps(p_3,q_3) \; p_1 \cdot \vareps(p_4,q_4) \,,  \\
T_{\gg,6} & = \bar{u}(p_2) \slashed{p}_3\slashed{p}_4 v(p_1) \; p_1 \cdot \vareps(p_3,q_3) \; p_2 \cdot \vareps(p_4,q_4) \,,  \\
T_{\gg,7} & = \bar{u}(p_2) \slashed{p}_3\slashed{p}_4 v(p_1) \; p_2 \cdot \vareps(p_3,q_3) \; p_1 \cdot \vareps(p_4,q_4) \,,  \\
T_{\gg,8} & = \bar{u}(p_2) \slashed{p}_3\slashed{p}_4 v(p_1) \; p_2 \cdot \vareps(p_3,q_3) \; p_2 \cdot \vareps(p_4,q_4) \,, 
\end{aligned}
\end{align}
and
\begin{align}
\label{eq:bbqqHtensorstructure}
\begin{aligned}
T_{\qq,1} & = \bar{u}(p_2) v(p_1) \; \bar{u}(p_4) \slashed{p}_1 v(p_3) \,,  \\
T_{\qq,2} & = \bar{u}(p_2) v(p_1) \; \bar{u}(p_4) \slashed{p}_2 v(p_3) \,,  \\
T_{\qq,3} & = \bar{u}(p_2) \slashed{p}_3\slashed{p}_4 v(p_1) \; \bar{u}(p_4) \slashed{p}_1 v(p_3) \,, \\
T_{\qq,4} & = \bar{u}(p_2) \slashed{p}_3\slashed{p}_4 v(p_1) \; \bar{u}(p_4) \slashed{p}_2 v(p_3) \,. 
\end{aligned}
\end{align}
Here, $\vareps(p_i, q_i)$ denotes the polarisation vector of the gluon with momentum $p_i$, defined with respect to the arbitrary reference vector $q_i$.
We choose $q_3 = p_4$ and $q_4 = p_3$.
We recall that the external momenta $p_i$ are taken to be four dimensional.

The form factors $\omega^{(L)}_{\gg/\qq,i}$ are obtained from the amplitudes by inverting Eqs.~\eqref{eq:bbggHtensordecomposition} and~\eqref{eq:bbqqHtensordecomposition},
\begin{subequations} \label{eq:form_factors}
\begin{align}
	\omega^{(L)}_{\gg,i} & = \sum_{j=1}^{8} \left( \Theta^{-1}_{\gg} \right)_{ij} \sum_{\mathrm{pol}} T^\dagger_{\gg,j} A^{(L)}_{\gg}  \,, \\
	\omega^{(L)}_{\qq,i} & = \sum_{j=1}^{4} \left( \Theta^{-1}_{\qq} \right)_{ij} \sum_{\mathrm{pol}} T^\dagger_{\qq,j} A^{(L)}_{\qq}  \,,
\end{align}
\end{subequations}
where the sums run over all polarisations, and
\begin{equation}
\Theta_{\gg,ij} = \sum_{\mathrm{pol}} T^\dagger_{\gg,i} T_{\gg,j} \qquad \mathrm{and} \qquad
\Theta_{\qq,ij} = \sum_{\mathrm{pol}} T^\dagger_{\qq,i} T_{\qq,j} \,.
\end{equation}
We use the following form of the gluon polarisation vector sum:
\begin{equation}
\label{eq:polarisationsum}
\sum_{\mathrm{pol}} \vareps^{*}_{\mu}(p_i,q_i) \, \vareps_{\nu}(p_i,q_i) =
-g_{\mu\nu} + \frac{ p_{i\mu}q_{i\nu} + q_{i\mu}p_{i\nu} }{ p_i \cdot q_i}, \qquad i=3,4 \,.
\end{equation}
To obtain the helicity amplitudes, we first specify the helicity states of the massless spinors and polarisation vectors appearing in the tensor structures specified in
Eq.~\eqref{eq:bbggHtensorstructure} for $\bbggH$,
\begin{align}
\label{eq:bbggHhelicitytensorstructure}
\begin{aligned}
T^{h_1 h_2 h_3 h_4}_{\gg,1} & = \bar{u}(p_2,h_2) v(p_1,h_1) \; p_1 \cdot \vareps(p_3,q_3,h_3) \; p_1 \cdot \vareps(p_4,q_4,h_4) \,,  \\
T^{h_1 h_2 h_3 h_4}_{\gg,2} & = \bar{u}(p_2,h_2) v(p_1,h_1) \; p_1 \cdot \vareps(p_3,q_3,h_3) \; p_2 \cdot \vareps(p_4,q_4,h_4) \,,  \\
T^{h_1 h_2 h_3 h_4}_{\gg,3} & = \bar{u}(p_2,h_2) v(p_1,h_1) \; p_2 \cdot \vareps(p_3,q_3,h_3) \; p_1 \cdot \vareps(p_4,q_4,h_4) \,,  \\
T^{h_1 h_2 h_3 h_4}_{\gg,4} & = \bar{u}(p_2,h_2) v(p_1,h_1) \; p_2 \cdot \vareps(p_3,q_3,h_3) \; p_2 \cdot \vareps(p_4,q_4,h_4) \,,  \\
T^{h_1 h_2 h_3 h_4}_{\gg,5} & = \bar{u}(p_2,h_2) \slashed{p}_3\slashed{p}_4 v(p_1,h_1) \; p_1 \cdot \vareps(p_3,q_3,h_3) \; p_1 \cdot \vareps(p_4,q_4,h_4) \,,  \\
T^{h_1 h_2 h_3 h_4}_{\gg,6} & = \bar{u}(p_2,h_2) \slashed{p}_3\slashed{p}_4 v(p_1,h_1) \; p_1 \cdot \vareps(p_3,q_3,h_3) \; p_2 \cdot \vareps(p_4,q_4,h_4) \,,  \\
T^{h_1 h_2 h_3 h_4}_{\gg,7} & = \bar{u}(p_2,h_2) \slashed{p}_3\slashed{p}_4 v(p_1,h_1) \; p_2 \cdot \vareps(p_3,q_3,h_3) \; p_1 \cdot \vareps(p_4,q_4,h_4) \,,  \\
T^{h_1 h_2 h_3 h_4}_{\gg,8} & = \bar{u}(p_2,h_2) \slashed{p}_3\slashed{p}_4 v(p_1,h_1) \; p_2 \cdot \vareps(p_3,q_3,h_3) \; p_2 \cdot \vareps(p_4,q_4,h_4) \,, 
\end{aligned}
\end{align}
and Eq.~\eqref{eq:bbqqHtensorstructure} for $\bbqqH$,
\begin{align}
\label{eq:bbqqHhelicitytensorstructure}
\begin{aligned}
T^{h_1 h_2 h_3 h_4}_{\qq,1} & = \bar{u}(p_2,h_2) v(p_1,h_1) \; \bar{u}(p_4,h_4) \slashed{p}_1 v(p_3,h_3) \,,  \\
T^{h_1 h_2 h_3 h_4}_{\qq,2} & = \bar{u}(p_2,h_2) v(p_1,h_1) \; \bar{u}(p_4,h_4) \slashed{p}_2 v(p_3,h_3) \,,  \\
T^{h_1 h_2 h_3 h_4}_{\qq,3} & = \bar{u}(p_2,h_2) \slashed{p}_3\slashed{p}_4 v(p_1,h_1) \; \bar{u}(p_4,h_4) \slashed{p}_1 v(p_3,h_3) \,, \\
T^{h_1 h_2 h_3 h_4}_{\qq,4} & = \bar{u}(p_2,h_2) \slashed{p}_3\slashed{p}_4 v(p_1,h_1) \; \bar{u}(p_4,h_4) \slashed{p}_2 v(p_3,h_3) \,. 
\end{aligned}
\end{align}
We use the following expressions in the spinor-helicity formalism for the quarks' wave functions $u$ and $v$, and for the gluons' polarisation vectors with a definite helicity state:
\begin{alignat}{3}
\label{eq:spinorhelicitydefinition}
\begin{aligned}
\bar{u}(p_i,+) & = [i | \,, \qquad
v(p_i,+) & = | i ] \,, \qquad 
\vareps^\mu(p_i,p_j,+) & = \frac{\spAB{j}{\gamma^\mu}{i}}{\sqrt{2}\spA{j}{i}}\,, \\
\bar{u}(p_i,-) & = \langle i | \,, \qquad
v(p_i,-) & = | i \rangle \,, \qquad
\vareps^\mu(p_i,p_j,-) & = \frac{\spAB{i}{\gamma^\mu}{j}}{\sqrt{2}\spB{i}{j}}\,,
\end{aligned}
\end{alignat}
for massless momenta $p_i$ and $p_j$.
Finally, the helicity amplitudes are obtained by multiplying the helicity tensor structures $T^{h_1 h_2 h_3 h_4}_{\gg/\qq,i}$ by the form factors $\omega^{(L)}_{\gg/\qq,i}$ as given in Eqs.~\eqref{eq:form_factors}:
\begin{subequations}
\label{eq:helicityamps}
\begin{align}
A^{(L),h_1 h_2 h_3 h_4}_{\gg} & = \sum_{i,j=1}^{8} T^{h_1 h_2 h_3 h_4}_{\gg,i} \, \left( \Theta^{-1}_{\gg} \right)_{ij} \, \sum_{\mathrm{pol}} T^\dagger_{\gg,j} A^{(L)}_{\gg} \,,  \\
A^{(L),h_1 h_2 h_3 h_4}_{\qq} & = \sum_{i,j=1}^{4} T^{h_1 h_2 h_3 h_4}_{\qq,i} \, \left( \Theta^{-1}_{\qq} \right)_{ij} \, \sum_{\mathrm{pol}} T^\dagger_{\qq,j} A^{(L)}_{\qq} \,.
\end{align}
\end{subequations}
We note that, at this stage, $\Theta_{\gg/\qq}$ and the coefficients of the Feynman integrals/special functions appearing in $\sum_{\mathrm{pol}} T^\dagger_{\gg/\qq,i} A^{(L)}_{\gg/\qq}$
are functions of the Mandelstam invariants $\vec{s}$ while, due to Eq.~\eqref{eq:spinorhelicitydefinition},
$T^{h_1 h_2 h_3 h_4}_{\gg/\qq,i}$ is written in terms of spinor products ($\spA{i}{j}$ and $\spB{i}{j}$).

\subsection{Analytic computation with finite-field arithmetic}
\label{sec:reduction}

Our workflow for the analytic computation of the helicity amplitudes up to two-loop level, to a large extent, 
makes use of the codebase developed in several previous amplitude calculations~\cite{Badger:2021nhg,Badger:2021imn,Badger:2022ncb,Badger:2024sqv}.
We start with the generation of the relevant Feynman diagrams by means of \textsc{Qgraf}~\cite{Nogueira:1991ex}, then perform the colour decomposition 
to separate colour from kinematics, and expand the amplitude in $N_c$ and $n_f$ in order to obtain the partial sub-amplitudes, following
Eqs.~\eqref{eq:bbggHcolourdecomposition}~--~\eqref{eq:bbggHNcNf2Ldecomposition} and 
Eqs.~\eqref{eq:bbqqHcolourdecomposition}~--~\eqref{eq:bbqqHNcNfdecomposition}.
For each partial sub-amplitude, we identify a minimal set of integral families with the maximum number of loop-momentum dependent propagators, i.e., 5 (8) propagators at one loop (two loops).
All diagrams, including those with fewer propagators, are mapped onto these integral families.
We then construct the contracted amplitude $\sum_{\mathrm{pol}} T^\dagger_{\gg/\qq,i} A^{(L)}_{\gg/\qq}$ 
by multiplying the amplitude with the conjugated tensor structures defined in
Eqs.~\eqref{eq:bbggHtensorstructure}~and~\eqref{eq:bbqqHtensorstructure}, and summing over both the external polarisation states and the spinor indices.
Lorentz contractions, Dirac $\gamma$-matrix manipulations and traces are handled with a combination of \textsc{Mathematica}
and \textsc{Form}~\cite{Kuipers:2012rf,Ruijl:2017dtg} scripts.
After these operations, the numerators of the contracted loop amplitude contain
scalar products involving loop and external momenta ($k_i \cdot k_j$, $k_i \cdot p_j$
and $p_i \cdot p_j$), which are then expressed in terms of the propagators of the integral family associated with the diagram under consideration.

The contracted loop amplitude, for a given colour structure and for given powers of $N_c$ and $n_f$, is now expressed as a linear combination of scalar Feynman integrals $\cI_j$,
\begin{equation}
	\label{eq:ampintegrand}
	\sum_{\mathrm{pol}} T^\dagger_{\gg/\qq,i} A^{(L)}_{\gg/\qq} = 
	\sum_{j} c^{(L)}_{\gg/\qq;ij}(\eps,\vec{s}) \; \cI_j \,.
\end{equation}
From this stage onwards, all the manipulations of the rational functions are performed through numerical evaluations over finite fields, and concatenated in a dataflow graph within the \textsc{FiniteFlow} framework.

\begin{figure}[t!]
  \begin{center}
    \includegraphics[width=0.55\textwidth]{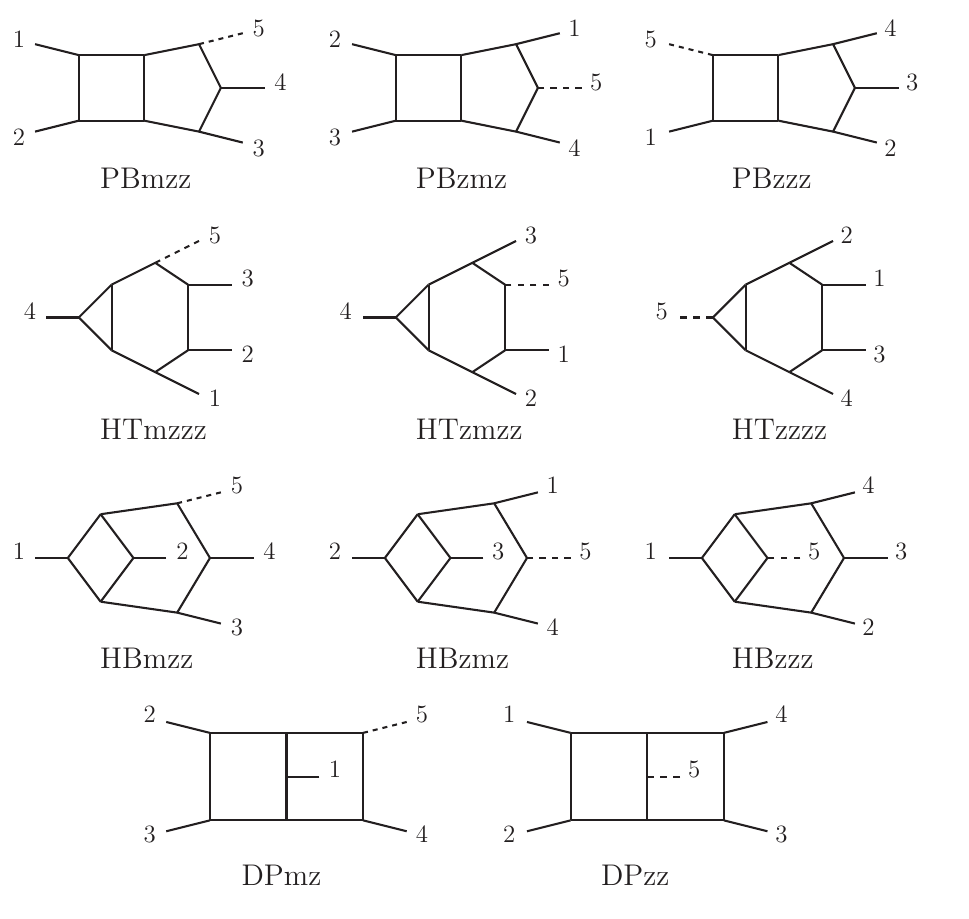}
  \end{center}
  \caption{Ordered two-loop integral families entering the IBP reduction:
            three pentagon-boxes (PB), three hexagon-triangles (HT), three hexagon-boxes (HB), and two double-pentagons (DP).}
  \label{fig:mainfam2L}
\end{figure}

The scalar Feynman integrals are reduced to the set of master integrals $\cG_j$ identified in Refs.~\cite{Abreu:2020jxa,Abreu:2021smk,Abreu:2023rco},
by means of integration-by-parts (IBP) reduction~\cite{Tkachov:1981wb,Chetyrkin:1981qh,Laporta:2000dsw},
\begin{equation}
	\label{eq:ampMIs}
	\sum_{\mathrm{pol}} T^\dagger_{\gg/\qq,i} A^{(L)}_{\gg/\qq} = 
	\sum_{j} d^{(L)}_{\gg/\qq;ij}(\eps,\vec{s}) \; \cG_j  \,.
\end{equation}
For this purpose, we generated optimised systems of IBP relations
by using \textsc{NeatIBP}~\cite{Wu:2023upw}, which employs the syzygy
method~\cite{Gluza:2010ws,Chen:2015lyz,Larsen:2015ped,Zhang:2016kfo,Bohm:2017qme,Bosma:2018mtf,Boehm:2020zig}, 
for each \emph{ordered} integral family.
We re-used the IBP relations generated to compute the two-loop amplitudes for $W\gamma\gamma$ production in Ref.~\cite{Badger:2024sqv}, and complemented them with few additional relations ---~also generated with \textsc{NeatIBP}~--- to reduce the scalar integrals which did not appear there.
We show the representative graphs of all ordered two-loop integral families in Fig.~\ref{fig:mainfam2L}.
The full set of integral families covering the entire amplitude can be obtained from the ordered ones by appropriately permuting the external momenta.
For each individual ordered integral family, then, the optimised IBP relations are loaded into \textsc{FiniteFlow} graphs and solved numerically over finite fields with Laporta's algorithm~\cite{Laporta:2000dsw}.
The solution to the IBP relations for the permuted families is obtained by evaluating the solution for the ordered families at permuted values of the kinematic variables.
To express the helicity amplitude in terms of a set of master integrals $\{\cG_j\}$ that is minimal across all integral families and permutations thereof, we obtained relations between these
with \textsc{LiteRed}~\cite{Lee:2012cn}. Since we use \emph{pure} bases of master integrals~\cite{Henn:2013pwa}, the relations among them do not depend on the kinematics.
We note the IBP reduction approach adopted here has been previously applied in a number of amplitude
computations~\cite{Badger:2021imn,Badger:2023mgf,Badger:2023xtl,Badger:2024sqv},
and a pedagogical illustration of the method can be found in Appendix~B of Ref.~\cite{Badger:2023xtl}.

We show in Table~\ref{tab:integrandIBPdata} the number of external-momentum permutations required to cover all the scalar Feynman integrals appearing in the full amplitude,
and the file size of IBP systems for each ordered two-loop integral family.\footnote{For each ordered integral family, the corresponding IBP relations are stored in an uncompressed text file in \textsc{Mathematica} format.}
In our framework, the evaluation time of the coefficients of the master integrals ($d^{(L)}_{\gg/\qq;ij}(\eps,\vec{s})$
in Eq.~\eqref{eq:ampMIs}) depends on both the number of IBP equations (which is correlated to the file size of IBP system)
and the number of external-momentum permutations appearing in each ordered integral family. The memory consumption, however, only depends on the size of
the IBP systems of the ordered integral families, making this approach suitable for calculations where a large number of external-momentum
permutations are required.

\renewcommand{\arraystretch}{1.5}
\begin{table}[]
\centering
\begin{tabular}{|c|c|c|c|c|c|c|}
\hline
   \multicolumn{1}{|c|}{}    & \multicolumn{1}{c|}{$A_{34}^{(2),N_c^2}$} & \multicolumn{1}{c|}{$A_{34}^{(2),1}$} & \multicolumn{1}{c|}{$A_{34}^{(2),1/N_c^2}$} 
	& \multicolumn{1}{c|}{$A_{\delta}^{(2),N_c}$} & \multicolumn{1}{c|}{$A_{\delta}^{(2),1/N_c}$} & \multirow{2}{*}{\makecell{\textsc{NeatIBP}\\file size}} \\\cline{1-6}
   \multicolumn{1}{|c|}{ \# of scalar integrals} & \multicolumn{1}{c|}{20905} & \multicolumn{1}{c|}{147235} & \multicolumn{1}{c|}{83731} & \multicolumn{1}{c|}{215394} & \multicolumn{1}{c|}{204375} &  \\\hline
	$N_\mathrm{perm}$(PBmzz) & 2 & 16 &  7 &  8 &  8 & 8MB   \\
	$N_\mathrm{perm}$(PBzmz) & 1 &  7 &  4 &  4 &  3 & 10MB  \\
	$N_\mathrm{perm}$(PBzzz) & 2 &  4 &  7 & 12 & 24 & 5.9MB \\
	\hline
	$N_\mathrm{perm}$(HTmzzz) & 2 &  8 & 12 & 24 & 24 & 1.9MB \\
	$N_\mathrm{perm}$(HTzmzz) & 2 & 14 &  8 & 24 & 20 & 3.8MB \\
	$N_\mathrm{perm}$(HTzzzz) & 1 &  4 &  4 & 12 & 12 & 4MB   \\
	\hline
	$N_\mathrm{perm}$(HBmzz) & 0 & 10 &  6 & 10 & 12 & 17MB  \\
	$N_\mathrm{perm}$(HBzmz) & 0 &  5 &  3 &  5 &  5 & 13MB  \\
	$N_\mathrm{perm}$(HBzzz) & 0 &  4 &  2 &  6 &  6 & 38MB  \\
	\hline
	$N_\mathrm{perm}$(DPmz)  & 0 & 18 &  8 & 16 & 24 & 71MB  \\
	$N_\mathrm{perm}$(DPzz)  & 0 &  4 &  1 &  2 &  2 & 376MB \\
\hline
\end{tabular}
	\caption{Number of scalar integrals appearing in the two-loop $\bbggH$ partial amplitudes at the various orders in $N_c$ without
	closed fermion loops. For each partial amplitude, we report the number of external leg permutations, $N_\mathrm{perm}$, 
	for each of the ordered integral families shown in Figure~\ref{fig:mainfam2L}, together with the file size of the corresponding IBP relations as produced by
	\textsc{NeatIBP}, which are stored in uncompressed text files as \textsc{Mathematica} expressions.}
\label{tab:integrandIBPdata}
\end{table}

Next, the master integrals are expressed, order by order in $\eps$, as $\mathbb{Q}$-linear combinations of monomials of pentagon functions and transcendental constants ($\i \pi$ and $\zeta_3$)~\cite{Chicherin:2021dyp,Abreu:2023rco},\footnote{The monomials contain also square roots
(see Ref.~\cite{Abreu:2023rco} for their definition), as we moved the square-root normalisations from the master integrals to their expression in terms of special functions.} denoted $m_i(f)$.
The rational coefficients multiplying the master integrals are instead Laurent-expanded around $\eps=0$ within the finite-field workflow~\cite{Peraro:2019svx}.
The $\eps$-expansion is truncated at $\cO(\eps^2)$ at one loop and $\cO(\eps^0)$ at two loops:
\begin{equation}
	\label{eq:amppfuncs}
	\sum_{\mathrm{pol}} T^\dagger_{\gg/\qq,i} A^{(L)}_{\gg/\qq} = 
	\sum_{p=-2L}^{-2L+4} \sum_{j} \eps^p \; e^{(L)}_{\gg/\qq;ij,p}(\vec{s}) \; m_{j,p}(f)  \; + \; \cO(\eps^{-2L+5}) \,,
	\quad L \in \{1,2\}.
\end{equation}

We then convert the contracted amplitude to the helicity amplitude
according to Eq.~\eqref{eq:helicityamps}. Our setup allows for the simultaneous computation of multiple helicity configurations for a given partial sub-amplitude. In this view, we define the following sets of helicity amplitudes for $\bbggH$ and $\bbqqH$,
\begin{subequations}
\label{eq:helicityampsets}
\begin{align}
	A^{(L),\vec{h}_{g}}_{\gg} 
	& = \begin{pmatrix}  
		A^{(L)}_{\gg}(1_{\bar{b}}^+,2_b^+,3_g^+,4_g^+,5_H) \\ 
		A^{(L)}_{\gg}(1_{\bar{b}}^+,2_b^+,3_g^+,4_g^-,5_H) \\ 
		A^{(L)}_{\gg}(1_{\bar{b}}^+,2_b^+,3_g^-,4_g^-,5_H) 
	\end{pmatrix} \,, \\
	A^{(L),\vec{h}^\prime_{g}}_{\gg} 
	& = \begin{pmatrix}  
		A^{(L)}_{\gg}(1_{\bar{b}}^+,2_b^+,3_g^+,4_g^+,5_H) \\ 
		A^{(L)}_{\gg}(1_{\bar{b}}^+,2_b^+,3_g^+,4_g^-,5_H) \\ 
		A^{(L)}_{\gg}(1_{\bar{b}}^-,2_b^-,3_g^+,4_g^+,5_H) 
	\end{pmatrix} \,, \\
	A^{(L),\vec{h}_q}_{\qq} 
	& = \begin{pmatrix}  
		A^{(L)}_{\qq}(1_{\bar{b}}^+,2_b^+,3_{\bar{q}}^+,4_q^-,5_H) 
	\end{pmatrix} \,,
\end{align}
\end{subequations}
as well as the following helicity tensor matrices,
\begin{subequations}
\label{eq:helicitytensormatrix}
\begin{align}
	\cT^{\vec{h}_{g}}_{\gg} 
	& = \begin{pmatrix}  
		T^{++++}_{\gg,1} & \cdots & T^{++++}_{\gg,8}\\ 
		T^{+++-}_{\gg,1} & \cdots & T^{+++-}_{\gg,8}\\ 
		T^{++--}_{\gg,1} & \cdots & T^{++--}_{\gg,8}
	\end{pmatrix} \,, \\
	\cT^{\vec{h}^\prime_{g}}_{\gg} 
	& = \begin{pmatrix}  
		T^{++++}_{\gg,1} & \cdots & T^{++++}_{\gg,8}\\ 
		T^{+++-}_{\gg,1} & \cdots & T^{+++-}_{\gg,8}\\ 
		T^{--++}_{\gg,1} & \cdots & T^{--++}_{\gg,8}
	\end{pmatrix} \,, \\
	\cT^{\vec{h}_{q}}_{\qq} 
	& = \begin{pmatrix}  
		T^{+++-}_{\qq,1} & \cdots & T^{+++-}_{\qq,8}
	\end{pmatrix} \,.
\end{align}
\end{subequations}
For $\bbggH$ we define a second set of helicity configurations ($\vec{h}^\prime_{g}$) in addition to the default one in Eq.~\eqref{eq:indephelgg} ($\vec{h}_{g}$) because
the $A^{(2),1}_{34}$ and $A^{(2),1/N_c^2}_{34}$ sub-leading colour partial amplitudes are moderately easier to compute with $\vec{h}^\prime_{g}$ than with $\vec{h}_{g}$.
We express the helicity tensor matrices 
$\cT^{\vec{h}_{g}}_{\gg}$, 
$\cT^{\vec{h}^\prime_{g}}_{\gg}$ and
$\cT^{\vec{h}_{q}}_{\qq}$ 
in terms of the momentum-twistor variables $\vec{x}$ defined in Section~\ref{sec:Kinematics}.

The helicity amplitudes for a given set of helicity configurations $\vec{H} \in \{\vec{h}_{g},\vec{h}^\prime_{g},\vec{h}_q\}$ are then given by
\begin{align}
\label{eq:helampfuncs0}
	A^{(L),\vec{H}}_{\gg/\qq;i} & =  \sum_{p=-2L}^{-2L+4} \, \sum_{jkl} \, \cT^{\vec{H}}_{\gg/\qq,ij}(\vec{x})  \, \left[ \Theta^{-1}_{\gg/\qq}(\vec{s}) \right]_{jk} \, 
					 \eps^p \; e^{(L)}_{\gg/\qq;kl,p}(\vec{s}) \; m_{l,p}(f) \; + \; \cO(\eps^{-2L+5}) \,, \\
\label{eq:helampfuncs}
				    & =  \sum_{p=-2L}^{-2L+4} \, \sum_{j} \, \eps^p \; g^{(L),\vec{H}}_{\gg/\qq;ij,p}(\vec{x}) \; m_{j,p}(f) \; + \; \cO(\eps^{-2L+5}) \,,
\end{align}
for $L \in \{1,2\}$. All operations on rational functions from Eq.~\eqref{eq:ampintegrand} to Eq.~\eqref{eq:helampfuncs} are performed numerically using finite field arithmetic within \textsc{FiniteFlow}.
Also the change of variables $\vec{s} \to \vec{x}$ required to go from Eq.~\eqref{eq:helampfuncs0} to Eq.~\eqref{eq:helampfuncs0} (given in Eq.~\eqref{eq:ampintegrand}) is implemented within the finite-field framework.

At this stage, therefore, we have constructed a numerical algorithm to evaluate the rational coefficients of the pentagon-function monomials in the helicity amplitudes
($g^{(L),\vec{H}}_{\gg/\qq;ij,p}(\vec{x})$ in Eq.~\eqref{eq:helampfuncs}) over finite fields. The remaining task is to perform the functional reconstruction of the
rational coefficients from their numerical finite-field samples.\footnote{We reconstructed the coefficients of the bare helicity amplitudes, rather than those of the finite remainders,
as the two representations have similar complexity in terms of the required number of sample points and evaluation~time.}
This task is however made challenging by the high polynomial degrees, which leads to the need to evaluate a large number of sample points, as well as by the slow evaluation time, which is due to large number of integral family permutations in the IBP reduction step and to the complexity of the integrands.
To alleviate this issue, we implement a number of optimisations to reduce the required number of sample points, following the strategy
introduced in Refs.~\cite{Badger:2021nhg,Badger:2021imn}.

\renewcommand{\arraystretch}{1.5}
\begin{table}[]
\centering
\begin{tabular}{|c|c|cc|cc|cc|}
\hline
	& $A_{34}^{(2),N_c^2}$ & \multicolumn{2}{c|}{$A_{34}^{(2),1}$} &  \multicolumn{2}{c|}{$A_{\delta}^{(2),N_c}$} 
	&  \multicolumn{2}{c|}{$A_{\delta}^{(2),1/N_c}$} \\
\hline
	helicity set & $\vec{h}_g$ & \multicolumn{2}{c|}{$\vec{h}^\prime_g$} &  \multicolumn{2}{c|}{$\vec{h}_g$} &  \multicolumn{2}{c|}{$\vec{h}_g$} \\
\hline
	$x_1 = 1$ & 133/132 & 176/175  & 176/175 & 265/269 & 265/269 & 214/207 & 214/207 \\
\hline
	\makecell{ansatz for \\linear \\ relations} & - & - & {\small $A_{34}^{(2),1/N_c^2}$} & - 
	& \makecell{\small $A_{34}^{(2),1}$ \\ 	\small $A_{43}^{(2),1}$ \\   \small $A_{34}^{(2),1/N_c^2}$ \\   \small $A_{43}^{(2),1/N_c^2}$ \\  \small $A_{\delta}^{(2),1/N_c}$} 
	& - &  
        \makecell{\small $A_{34}^{(2),1}$ \\ \small $A_{43}^{(2),1}$ \\  \small $A_{34}^{(2),1/N_c^2}$ \\   \small $A_{43}^{(2),1/N_c^2}$  }  \\ 
\hline
	\makecell{linear\\relations}             & 133/132 & 176/175  & 110/109 & 233/233 & 121/115 & 141/140 & 135/130 \\
\hline
	\makecell{number of\\independent \\ coefficients}       & 162     & 572       & 216     & 582     & 151     & 581     & 258 \\
\hline
	\makecell{denominator\\matching}         & 133/0   & 176/0     & 110/0   & 233/0   & 119/0   & 141/0   & 135/0 \\
\hline
	\makecell{univariate\\partial\\fraction} & 32/25   & 38/33     & 25/22   & 36/33   & 36/33   & 32/28   & 32/28 \\
\hline
	\makecell{factor\\matching}              & 28/0    & 38/0      & 24/0    & 36/0    & 36/0    & 32/0    & 32/0 \\
\hline
	\makecell{number of\\ sample points}             & 16711   & 47608     & 10382   & 38221   & 37002   & 21150   & 20337\\
\hline
	\makecell{evaluation\\time per point}             & $t_0$   & $100 t_0$ & $67 t_0$    & $106 t_0$    & $57 t_0$    & $74 t_0$    & $62 t_0$ \\
\hline
\end{tabular}
\caption{Functional reconstruction data for the three most complicated two-loop subleading colour 
	$\bbggH$ amplitudes, as well as for the most complicated two-loop leading colour amplitude, for comparison. We report the maximal numerator/denominator polynomial degree 
	of the rational coefficients appearing in the bare amplitude ($g^{(L),\vec{H}}_{\gg;ij,p}$ in Eq.~\eqref{eq:helampfuncs}) 
	at each stage of the reconstruction strategy discussed in the text.
	We also show the number of independent rational coefficients, the number of sample points required to fully reconstruct the coefficients analytically,
	and the evaluation time for each finite-field point relative to that of the leading colour amplitude.
	The fourth row lists the partial sub-amplitudes used to construct the ansatz for the rational coefficients (see Eq.~\eqref{eq:linrels}), with - meaning that no ansatz is made.
	}
\label{tab:reconstructiondata}
\end{table}

The first, trivial, step of optimisation is to set $x_1 = 1$. Reinstating the $x_1$ dependence after the reconstruction is straightforward:
$x_1$ is in fact the only dimensionful quantity in our momentum-twistor parametrisation,
and therefore it appears only as a global prefactor of the amplitude. Additionally, as the pentagon-function monomials $m_i(f)$ contain also square roots, we introduce powers of $x_1$ in the monomials in order to keep them dimensionless.
Secondly, we search for linear relations among the rational coefficients appearing in Eq.~\eqref{eq:helampfuncs} numerically over finite fields, and only reconstruct the linearly independent ones, 
which are chosen so as to minimise the polynomial degrees. 
This procedure can be enhanced by including in the linear relations an ansatz for the rational coefficients.
We construct this ansatz out of the rational coefficients of the partial sub-amplitudes which have already been reconstructed. 
In practice, when reconstructing the list of coefficients $\{g^{(L),\vec{H}}_{\gg/\qq;ij,p}(\vec{x}) \}$,
we first determine the linear relations
\begin{equation} \label{eq:linrels}
	\sum_{ijp} a_{ijp} \; g^{(L),\vec{H}}_{\gg/\qq;ij,p}(\vec{x}) + \sum_{ijp} b_{ijp} \; \tilde{g}^{(L),\vec{H}}_{\gg/\qq;ij,p}(\vec{x}) \; = \; 0\,,
\end{equation}
where $a_{ijp}, b_{ijp} \in \mathbb{Q}$, and $\{\tilde{g}^{(L),\vec{H}}_{\gg/\qq;ij,p}(\vec{x})\}$ are the rational coefficients of partial sub-amplitudes which have already been computed.
We then solve the relations in Eq.~\eqref{eq:linrels}, preferring the known $\{\tilde{g}^{(L),\vec{H}}_{\gg/\qq;ij,p}(\vec{x}) \}$ over the unknown $\{g^{(L),\vec{H}}_{\gg/\qq;ij,p}(\vec{x}) \}$, and the simpler $\{g^{(L),\vec{H}}_{\gg/\qq;ij,p}(\vec{x}) \}$ over the more complicated ones.
Once we have an independent set of rational coefficients, we determine the analytic form of their denominators by matching them onto ans\"atze built 
out of the symbol letters~\cite{Abreu:2023rco} 
and a collection of simple spinor products on a univariate phase-space slice. 
Using this information and the polynomial degrees of the unknown numerators, we perform an on-the-fly univariate partial fraction decomposition with respect to $x_5$ of the independent rational coefficients.
This is also achieved by writing an ansatz for the partial fraction decomposition, and reconstructing the coefficients in the ansatz, which are functions of $(x_2, x_3, x_4, x_6)$.
Finally, we perform another factor matching on a univariate slice, enlarging the ansatz so as to include also the spurious denominator factors introduced by the univariate partial fraction decomposition~\cite{Badger:2021imn}.
The remaining numerators of the coefficients of the partial fraction decompositions are reconstructed using \textsc{FiniteFlow}'s built-in functional reconstruction algorithm~\cite{Peraro:2016wsq}.

In Table~\ref{tab:reconstructiondata}, we show the maximum polynomial degrees in the numerator and denominator of the 
rational coefficients of the pentagon-function monomials at each stage
of the reconstruction strategy outlined above, for the three most complicated two-loop 
subleading colour $\bbggH$ amplitudes, as well as for the most complicated two-loop leading colour amplitude, for comparison.
The rest of the $\bbggH$ amplitudes and all of the $\bbqqH$ amplitudes are rather straightforward to reconstruct, and hence we do not present their
reconstruction data. We observe in Table~\ref{tab:reconstructiondata} that, prior to performing any optimisation, the maximum polynomial degree 
of the rational coefficients in the sub-leading colour amplitudes is significantly higher than in the leading colour case.
For $A_{\delta}^{(2),N_c}$ and $A_{\delta}^{(2),1/N_c}$, solving the linear relations 
without any ansatz leads to a drop in the maximum polynomial degree, while the latter stays the same for $A^{(2),1}_{34}$.
Using an ansatz from other partial amplitudes decreases the maximum polynomial degrees quite significantly for all three subleading
colour amplitudes. At the final optimisation stage, the numbers of required sample points are also smaller when the linear relations include coefficient ans\"atze, although the difference for $A_{\delta}^{(2),N_c}$ and $A_{\delta}^{(2),1/N_c}$ is not significant. 
The reduction in the number of sample points is however  accompanied by a speed-up in the finite-field evaluation when the coefficient ansatz is included in the linear relations.
To understand this, we recall that sampling the coefficients of the partial fraction decomposition involves the solution of a linear system of equations, which increases the evaluation time by a factor which is roughly equal to the number of terms in the ans\"atze for the partial fraction decomposition~\cite{Badger:2021imn}.
From the last row of Table~\ref{tab:reconstructiondata}, we thus see that, although the coefficient ansatz does not reduce the maximum degrees for $A_{\delta}^{(2),N_c}$ and $A_{\delta}^{(2),1/N_c}$, it does make the finite-field sampling faster by shortening the partial fractioning ans\"atze.
Finally, we see a slow-down of at least 50 times from the most complicated leading colour partial amplitude to the most complicated subleading colour one.
This can be attributed to: (i) the larger number of integral family permutations appearing in the 
IBP reduction, (ii) the more complicated IBP systems that contribute, most notably DPmz and DPzz families, and (iii) the larger size of the integrands, as represented by the number of scalar Feynman integrals shown in Table~\ref{tab:integrandIBPdata}.
Additionally, the memory usage for the subleading colour amplitudes presented in Table~\ref{tab:reconstructiondata} is 
roughly 3 times higher than for the most complicated leading colour amplitude.

With the functional reconstruction strategy discussed above we are finally able to obtain the analytic expression 
of the bare helicity amplitudes for the helicity configurations listed in Eq.~\eqref{eq:helicityampsets}.
In order to present the $\bbggH$ finite remainders for the independent helicity configurations in Eq.~\eqref{eq:indephelgg}, we convert the $A^{(2),1;--++}_{34}$ and $A^{(2),1/N_c^2;--++}_{34}$ bare amplitudes into $A^{(2),1;++--}_{34}$ and $A^{(2),1/N_c^2;++--}_{34}$.
We then extract the finite remainders from the bare amplitudes according to Eq.~\eqref{eq:finiteremainder}.
The helicity-dependent finite remainders now take the form
\begin{equation}
	\label{eq:finiteremainder2}
	F^{(L),h_1 h_2 h_3 h_4 }_{\gg/\qq}  = \sum_{p=0}^{-2L+4}  \sum_{i} \, \eps^p \, r^{(L),h_1 h_2 h_3 h_4}_{\gg/\qq;i,p}(\vec{x}) \; m_{i,p}(f)  \; + \;  \cO(\eps^{-2L+5}) \,,
	\quad L \in \{1,2\}.
\end{equation}

In order to optimise the expressions, we write all finite remainders for each loop order and helicity configuration in terms of a global set of linearly independent coefficients, chosen so as to minimise the polynomial degrees.
We then perform a multivariate partial fraction decomposition of the independent coefficients by means of \textsc{pfd-parallel}~\cite{Bendle:2021ueg}.
Following the strategy outlined in Ref.~\cite{Badger:2024sqv}, we decompose each addend in the univariate partial-fractioned representation rather than the complete coefficients. 
For the terms whose univariate partial-fractioned representation was more compact than the multivariate one, we kept the former.
Note that the univariate partial fraction decomposition can introduce spurious
factors in the denominators that do not correspond to physical singularities.
In principle, a multivariate partial fraction decomposition can eliminate such
spurious terms.  However, the computations required to do this can be extremely
expensive and so we take a practical approach in which some spurious denominator
factors remain, yet we see significant simplifications in the expressions.
Numerical tests of this representation did not show any significant instabilities.
Finally, we write the addends of the partial-fractioned coefficients in terms of independent irreducible factors, collected across each set of independent coefficients.

\subsection{Validation}

In this section we list the checks we have performed to validate our results.

\begin{itemize}

	\item {\bf Comparison against direct helicity amplitude calculation}

		We set up the direct computation of the helicity amplitudes as an alternative to the tensor-decomposition method discussed in Section~\ref{sec:tensordecomposition}. In this approach, the helicity-dependent loop numerators are constructed directly from the Feynman diagrams, and further expressed as linear combinations of scalar Feynman integrals.
		The subsequent steps are the same as described in Section~\ref{sec:reduction}.
		More details on the direct helicity amplitude approach can be found in 
		Refs.~\cite{Hartanto:2019uvl,Badger:2021owl,Badger:2021imn,Badger:2021ega,Badger:2023mgf,Badger:2023xtl}.
		We cross-checked the two approaches at the level of the bare amplitudes in the form of Eq.~\eqref{eq:helampfuncs} for each independent helicity configuration by evaluating the rational coefficients at a non-physical, numerical phase-space point, while leaving pentagon functions, transcendental constants and square roots symbolic.

	\item {\bf Ward-identity test}

		We check that the $\bbggH$ amplitude vanishes upon replacing one of the gluon polarisation vectors with its momentum for each independent helicity configuration.
		In our test we replace $\vareps^\mu(p_4) \rightarrow p_4$. In order to perform this test with the computational setup described in
		Section~\ref{sec:amplitude_computation}, we first modify the
		tensor decomposition in Eq.~\eqref{eq:bbggHtensordecomposition} to
		\begin{align}
			A^{(L)}_{\gg}(\vareps(p_4) \rightarrow p_4) & = \sum_{i=1}^{4} \tilde{T}_{\gg,i} \, \tilde{\omega}^{(L)}_{\gg,i} \,,  
		\end{align}
		where
		\begin{align}
		\label{eq:bbggHtensorstructureGaugeCheck}
		\begin{aligned}
			\tilde{T}_{\gg,1} & = \bar{u}(p_2) v(p_1) \; p_1 \cdot \vareps(p_3,q_3)  \,,  \\
			\tilde{T}_{\gg,2} & = \bar{u}(p_2) v(p_1) \; p_2 \cdot \vareps(p_3,q_3)  \,,  \\
			\tilde{T}_{\gg,3} & = \bar{u}(p_2) \slashed{p}_3\slashed{p}_4 v(p_1) \; p_1 \cdot \vareps(p_3,q_3)  \,,  \\
			\tilde{T}_{\gg,4} & = \bar{u}(p_2) \slashed{p}_3\slashed{p}_4 v(p_1) \; p_2 \cdot \vareps(p_3,q_3)  \,.
		\end{aligned}
		\end{align}
		As in Section.~\ref{sec:tensordecomposition}, we choose $q_3=p_4$ and use the gluon-polarisation vector sum specified in Eq.~\eqref{eq:polarisationsum} for $\vareps(p_3)$.
		The amplitude vanishes already upon contraction with the conjugated tensor structures given in Eq.~\eqref{eq:bbggHtensorstructureGaugeCheck},
		$\sum_{\mathrm{pol}} \tilde{T}^{\dagger}_{\gg,i}A^{(L)}_{\gg}(\vareps(p_4) \rightarrow p_4)$, at the level of the master-integral representation in Eq~\eqref{eq:ampMIs}.
		This check is done using a non-physical, numerical phase-space point for the master-integral coefficients while leaving the master integrals symbolic.
 		The fact that the contracted amplitude vanishes in this Ward-identity check implies that the helicity amplitude vanishes as well. Additionally, we perform the Ward-identity check
		using the direct helicity amplitude setup, where the replacement  $\vareps^\mu(p_4) \rightarrow p_4$ is done directly in the construction of the helicity-dependent loop numerators.

	\item {\bf Pole cancellation}

		The successful computation of the finite remainders following the prescription of Section~\ref{sec:UVIRsubtraction} demonstrates that the $\eps$ poles are as expected from UV renormalisation and IR factorisation.

	\item {\bf $\mu$ dependence of the finite remainder}

		We validate the $\mu$-restoring terms for the finite remainders given in Section~\ref{sec:mudependence} by evaluating numerically the finite remainders at two phase-space points, $\vec{s}_A$ and $\vec{s}_B$, related by an arbitrary rescaling factor $a$, as
		\begin{subequations}
		\begin{align}
			\vec{s}_A & = \lbrace s_{12}, s_{23}, s_{34}, s_{45}, s_{15}, m_H^2 \rbrace \,, \\
			\vec{s}_B & = \frac{\vec{s}_A}{a} \,. 
		\end{align}
		\end{subequations}
		In terms of momentum-twistor variables, they correspond to
		\begin{subequations}
		\begin{align}
			\vec{x}_A & = \lbrace x_{1}, x_{2}, x_{3}, x_{4}, x_{5}, x_6 \rbrace \,, \\
			\vec{x}_B & = \left\lbrace \frac{x_{1}}{a}, x_{2}, x_{3}, x_{4}, x_{5}, x_6 \right\rbrace \,. \\
		\end{align}
		\end{subequations}
		We then verify that the following scaling relation holds,
		\begin{equation}
			\frac{F^{(L)}(\vec{x}_B,\mu=1)}{F^{(0)}(\vec{x}_B)} = \frac{F^{(L)}(\vec{x}_A,\mu=a)}{F^{(0)}(\vec{x}_A)} \,,
		\end{equation}
		for all one- and two-loop helicity finite remainders.

	\item {\bf Comparison against \textsc{OpenLoops} at tree level and one loop}

		We cross-checked the tree-level and one-loop hard functions against numerical results obtained from
        \textsc{OpenLoops}~\cite{Buccioni:2019sur}
		for both $\bbggH$ and $\bbqqH$. We made use of \textsc{OpenLoops}' \verb=pphbb= process library and
        set $n_f = 5$ to obtain the results in the 5FS.

\end{itemize}

\section{Results}
\label{sec:results}

We obtained analytic expressions for the two-loop five-point amplitudes contributing to the double-virtual corrections to $\bbh$ production at the LHC, taking into account all colour contributions.
The analytic form of the finite remainders and the corresponding $\eps$-pole terms (containing UV and IR singularities) up to two-loop level are provided in the ancillary files~\cite{zenodo}
for the independent partial amplitudes and helicity configurations.
We provide \textsc{Mathematica} scripts to evaluate numerically the bare amplitudes and the hard functions. 
In addition, the finite remainders and the evaluation of hard functions have been implemented in a \textsc{C++} library which is also included in the ancillary files~\cite{zenodo}.
We describe in detail the content of \textsc{Mathematica} ancillary files accompanying this paper in Appendix~\ref{app:ancillaryfiles}.
The content and usage instruction of the \textsc{C++} library is described in the \verb=anc_cpp/README= in the ancillary files~\cite{zenodo}.
In this section, we present benchmark numerical evaluations of the hard functions for all channels relevant for the $pp\to \bbh$ scattering process, describe the implementation of the finite remainders in a \textsc{C++} library and discuss certain analytic properties related to the pentagon functions.

\subsection{Benchmark numerical evaluation of the hard functions}
We evaluate the hard functions for all the partonic scattering channels contributing to $pp\to\bbh$:
\begin{subequations}
\label{eq:bbHchannels}
\begin{align}
	gg\to \bar{b}bH                   &:\qquad  g(-p_3) + g(-p_4) \to \bar{b}(p_1) + b(p_2) + H(p_5) \,, \\
	q\bar{q}\to \bar{b}bH             &:\qquad q(-p_3) + \bar{q}(-p_4) \to \bar{b}(p_1) + b(p_2) + H(p_5) \,, \\
	\bar{q}q\to \bar{b}bH             &:\qquad \bar{q}(-p_3) + q(-p_4) \to \bar{b}(p_1) + b(p_2) + H(p_5) \,, \\
	b\bar{b}\to \bar{b}bH             &:\qquad b(-p_3) + \bar{b}(-p_4) \to \bar{b}(p_1) + b(p_2) + H(p_5) \,, \\
	\bar{b}b\to \bar{b}bH             &:\qquad \bar{b}(-p_3) + b(-p_4)  \to \bar{b}(p_1) + b(p_2) + H(p_5) \,, \\
	bb\to bbH                         &:\qquad b(-p_3) + b(-p_4) \to b(p_1) + b(p_2) + H(p_5) \,, \\
	\bar{b}\bar{b}\to \bar{b}\bar{b}H &:\qquad \bar{b}(-p_3) + \bar{b}(-p_4) \to \bar{b}(p_1) + \bar{b}(p_2) + H(p_5) \,.
\end{align}
\end{subequations}
For this purpose, we need to apply permutations of the external momenta
and/or parity conjugation to the independent partial finite remainders  which we computed analytically in order to obtain the full set of partial finite remainders and helicity configurations.
These operations are implemented differently on the pentagon functions and on their rational coefficients, as we describe below.
\begin{itemize}

\item For the rational coefficients of the pentagon-function monomials ($r^{(L),h_1 h_2 h_3 h_4}_{\gg/\qq;j}(\vec{x})$ in Eq.~\eqref{eq:finiteremainder2}),
the permutation of the external momenta is achieved
by applying the permutation on the RHS of Eq.~\eqref{eq:momtwistor5pt} before evaluating $\vec{s}$ and $\tr_{\pm}$ numerically.
This yields the values of the momentum-twistor variables at the permuted phase-space point.
Similarly, in the case of parity conjugation, we exchange $\tr_+ \leftrightarrow \tr_-$ on the RHS of Eq.~\eqref{eq:momtwistor5pt}.
We stress that these operations can only be performed on momentum-twistor expressions which are free of spinor phases.
The rational coefficients must therefore be divided by the appropriate spinor-phase factor, given in Eqs.~\eqref{eq:bbggHphase}~and~\eqref{eq:bbqqHphase}, expressed in terms of momentum-twistor variables.
The spinor-phase factor is then reinstated after applying the permutation
and/or parity conjugation on its spinor expression ---~which retains the phase information~--- and converting it to momentum-twistor variables.
We refer to Appendix~C of Ref.~\cite{Badger:2023mgf} for a detailed discussion.

\item The pentagon functions are evaluated numerically with the library \textsc{PentagonFunctions++}~\cite{PentagonFunctions:cpp}.
The evaluation for permutations of the external momenta can be obtained by expressing the permuted pentagon functions in terms of the unpermuted ones, so that the pentagon functions need to be evaluated only for 
the unpermuted external momenta. To derive such relations, the pentagon functions are first rewritten in terms of
master-integral components, which are then permuted and expressed back
in terms of unpermuted pentagon functions. The pentagon-function permutation rules are provided in the ancillary files.
Note that the library \textsc{PentagonFunctions++} requires the input phase-space point to be in a specific $2\to 3$ channel.
The momenta used in this work ($p_i$) are related to the ones of Ref.~\cite{Abreu:2023rco} ($q_i$)~by
\begin{equation}
(q_1,q_2,q_3,q_4,q_5) = (p_5,p_4,p_3,p_2,p_1) \,,
\end{equation}
and we relabel the Mandelstam invariants of Ref.~\cite{Abreu:2023rco} by $t_{ij} = (q_i+q_j)^2$, to distinguish them from ours ($s_{ij}$).
Then, \textsc{PentagonFunctions++} works in the $t_{45}$ channel, while the physical region relevant for our amplitudes is the $t_{23}$ ($s_{34}$) channel. In order to evaluate the pentagon functions in the $t_{23}$ channel, we evaluate them at a permuted phase-space point which belongs to the $t_{45}$ channel, $(q_1,q_4,q_5,q_2,q_3)$, and apply the transformation which expresses the pentagon functions with $(q_1,q_2,q_3,q_4,q_5)$ ordering in terms of those with $(q_1,q_4,q_5,q_2,q_3)$.

\end{itemize}

\renewcommand{\arraystretch}{1.5}
\begin{table}
    \centering
    \begin{tabular}{|c|ccc|}
    \hline
    $\bbggH$ & $\cH^{(0)}$ [GeV$^{-2}$] & $\cH^{(1)}/\cH^{(0)}$ & $\cH^{(2)}/\cH^{(0)}$  \\
    \hline
    $gg \to \bar{b}bH$ & $5.9300960 \cdot 10^{-9}$ & -0.0074420656  & 0.083709468  \\
    \hline
    \hline
    $\bbqqH$ & $\cH^{(0)}$ [GeV$^{-2}$] & $\cH^{(1)}/\cH^{(0)}$ & $\cH^{(2)}/\cH^{(0)}$  \\
    \hline
    $q\bar{q} \to \bar{b}bH$ & $2.6534235 \cdot 10^{-10}$ & -0.39272804  & 0.10382080 \\
    $\bar{q}q \to \bar{b}bH$ & $2.6534235 \cdot 10^{-10}$ & -0.049218183 & 0.10470723 \\
    \hline
    \hline
    $\bbbbH$ & $\cH^{(0)}$ [GeV$^{-2}$] & $\cH^{(1)}/\cH^{(0)}$ & $\cH^{(2)}/\cH^{(0)}$  \\
    \hline
    $b\bar{b} \to \bar{b}bH$             & $1.9980726 \cdot 10^{-9}$ & -0.0029434331 & 0.013659790 \\
    $\bar{b}b \to \bar{b}bH$             & $2.2951653 \cdot 10^{-8}$ &  0.26157437   & 0.046471305 \\ 
    $bb \to bbH$                         & $1.0166396 \cdot 10^{-8}$ &  0.32618581   & 0.068890688 \\
    $\bar{b}\bar{b} \to \bar{b}\bar{b}H$ & $1.0166396 \cdot 10^{-8}$ &  0.32618581   & 0.068890688 \\
    \hline
    \end{tabular}
	\caption{Numerical values of the hard functions defined in Eqs.~\eqref{eq:hardfunctions} for the partonic scattering channels listed in
	Eqs.~\eqref{eq:bbHchannels} using the phase-space point and input parameters given in Eqs.~\eqref{eq:PSpoint}~and~\eqref{eq:inputparameters}.}
    \label{tab:num-hardfunction}
\end{table}

In Table~\ref{tab:num-hardfunction}, we show benchmark numerical values for the tree-level, one-loop and two-loop hard functions
using a physical phase-space point associated with the $pp \to b\bar{b}H$ scattering process,
\begin{align}
\label{eq:PSpoint}
\begin{aligned}
p_1 & = (457.490386921      ,  -279.844382235      ,  -140.455057318      ,  -333.552024023) \,, \\
p_2 & = (403.008270188      ,   228.438854552      ,   143.939260340      ,   299.186973151) \,, \\
p_3 & = (-500,  0,  0,  -500)\,, \\
p_4 & = (-500,  0,  0,   500)\,, \\
p_5 & = (139.501342891      ,   51.4055276828      ,  -3.48420302179      ,   34.3650508716)\,, 
\end{aligned}
\end{align}
with the following values of the input parameters:
\begin{equation}
	\label{eq:inputparameters}
	\mu = 173.2 \,, \qquad \alpha_s = 0.118 \,, \qquad y_b = 0.0192916457706 \,.
\end{equation}
The momenta $p_i$ and the renormalisation scale $\mu$ are given in the units of GeV.
Benchmark numerical results for the bare helicity amplitudes separately for all partial amplitudes and ($N_c$, $n_f$) contributions are provided in the ancillary files.

\subsection{Numerical implementation}

We implemented all finite remainders in a \textsc{C++} library to yield the IR and UV finite squared matrix elements after colour and helicity sums as needed for differential cross sections. To this end, we employ the \textsc{PentagonFunctions++}~\cite{Chicherin:2020oor} library and the \textsc{qd} library~\cite{QDlib} whenever higher numerical precision is needed. The implementation has been cross-checked against the independent \textsc{Mathematica} implementation used to derive the benchmark values in Table~\ref{tab:num-hardfunction}, for a couple of phase-space points. To ensure the numerical stability of the matrix elements for a given phase space point $p$ and renormalisation scale $\mu$, we employ the following rescue strategy (similar to Ref.~\cite{Badger:2023mgf}) to successively increase the numerical precision if needed:

\begin{enumerate}
\item[1)] The matrix element is evaluated in standard double precision for the rational and transcendental (i.e., the pentagon functions) parts 
	at the given phase-space point as well as in a rescaled point $p'= \{ a \times p_i \}_{i=1}^5$, $\mu^{\prime} = a \times \mu$. Here $a$ is a numerical constant (in practice, we use $a=10$).

\item[2)] We exploit the scaling properties of the matrix element to compute the relative difference between those two points
  \begin{equation}
    r^{\text{diff}} = \frac{|\cH^{(2)}(p,\mu) - a^2 \, \cH^{(2)}(p', \mu^{\prime})|}{|\cH^{(2)}(p,\mu) + a^2 \, \cH^{(2)}(p', \mu^{\prime})|}\;.
  \end{equation}
  From this quantity, we define the estimate of the number of correct digits as
  \begin{equation}
    n^{\text{cor}} = -\log_{10}(r^{\text{diff}}) \;.
  \end{equation}

\item[3)] We demand at least four digits, i.e., $n^{\text{cor}} > 4$, for a phase space point to be accepted. If the estimated precision is insufficient, we restart from 1) with a higher numerical precision.
\end{enumerate}

We implement the following four precision levels:
\begin{itemize}
  \item sd/sd: the rational and transcendental parts are evaluated in standard double-precision.
  \item dd/sd: the rational part is evaluated at double-double precision ($\approx 32$ digits) while the transcendental part is kept at double precision.
  \item qd/sd: same as before but with quad-double precision ($\approx 64$ digits).
  \item qd/dd: same as previous but with the transcendental part in double-double precision.
\end{itemize}

\begin{figure}[t!]
  \begin{center}
    \includegraphics[width=\textwidth]{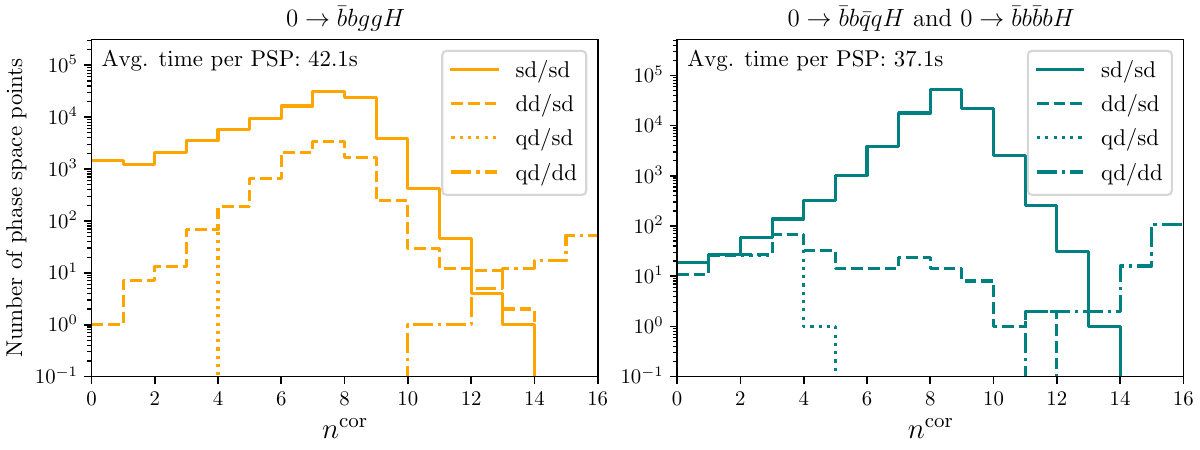}
  \end{center}
\caption{Estimated numerical precision of $\cH^{(2)}(p,\mu)$ evaluated for phase-space points arising during Monte Carlo integration given in terms of number of estimated correct digits. The results are separated according to the channels ($q\bar{q}/b\bar{b}/bb/\bar{b}\bar{b}$ and $gg$). Only points that fail the $n^{\text{cor}} > 4$ requirement are re-evaluated at the next higher numerical precision implemented. The estimated evaluation times contain the re-evaluations and are also averaged over the higher precision evaluations, and therefore represent the realistic times per phase-space point during Monte Carlo integration. In total 100000 points have been evaluated for each channel.}
  \label{fig:stability}
\end{figure}

To test the stability in a realistic example, we define a simple $b\bar{b}H$ phase space by requiring two anti-$k_T$ ($R=0.4$) $b$ jets with a transverse momentum of at least $25$ GeV and an absolute rapidity of at most $2.5$. We generate random points in this phase space and evaluate the two-loop finite remainders while keeping track of the number of correct digits achieved. In Figure~\ref{fig:stability}, we present the resulting distribution of correct digits organised in two sets of initial state channels $q\bar{q}/b\bar{b}/bb/\bar{b}\bar{b}$ and $gg$. Most of the points reach the stability goal with double-precision arithmetic. In the case of four-quark amplitudes, only $2-3$ permille of the points require a re-evaluation in higher precision arithmetic. These points appear to suffer mainly from the numerical accuracy of the transcendental part, as higher precision evaluation of the rational part does not improve the number of accurate digits. One permille of the points must be evaluated using double-double precision for the pentagon functions. For the two-quark two-gluon amplitudes, about 8 percent do not reach the accuracy goal with double-precision arithmetic. However, most unstable points can be rescued using higher precision rationals and, similarly to the quark-only amplitudes, only about one permille requires higher precision for the transcendental parts.

We find an average evaluation time of about 40 seconds for the amplitudes on a \textit{Intel(R) Xeon(R) Silver 4116 CPU @ 2.10GHz} CPU. The time estimate contains the re-evaluation with the rescaled points to assess the numerical stability and potential higher-precision evaluations. These times are sufficiently short to ensure that the evaluation of the two-loop contribution to the physical cross sections is not the bottleneck of a next-to-next-to-leading order computation.

In addition to the stability analysis we computed the double virtual contribution\footnote{We use the same conventions of the IR subtraction as in Ref.~\cite{Badger:2023mgf}.} to the total cross section for the above phase space and estimate it to be approximately 10\% with respect to the leading-order cross-section. Furthermore, we computed the contribution in the leading-colour approximation and estimate the effect of sub-leading colour terms to be about $\approx 2\%$ with respect to leading-order.

\subsection{Analytic properties}

We confirm the observations made in Ref.~\cite{Badger:2024sqv} regarding the analytic structure of the amplitudes, and refer to Section~4.2 thereof for a detailed discussion.
Here we content ourselves with recalling the most important features.
In particular, we observe that the symbol letter $\sqrt{\Delta_5}$ drops out of the finite remainders, while the six letters $\sqrt{\Sigma_5^{(i)}}$ ($i=1,\ldots,6$) are absent already from the bare amplitudes (see Ref.~\cite{Abreu:2023rco} for the definition of $\Delta_5$ and $\Sigma_5^{(i)}$, which are degree-4 polynomials in the Mandelstam invariants $\vec{s}$).
Thanks to the criteria adopted in the definition of the pentagon functions~\cite{Chicherin:2021dyp,Abreu:2023rco}, these letters are by construction isolated into minimal subsets of pentagon functions, and their cancellation is thus made manifest by the cancellation of the corresponding sets of pentagon functions.
This makes the analytic structure of the expressions more transparent, and the numerical evaluation more efficient.
We stress in particular that all pentagon functions which diverge in the physical phase space are absent, whereas all pentagon functions that do not involve the letters $\sqrt{\Delta_5}$ and $\sqrt{\Sigma_5^{(i)}}$ are present.

\section{Conclusion}
\label{sec:conclusion}

In this work we have presented the two-loop five-particle amplitudes contributing to $pp \to b\bar{b}H$ production at the LHC in the five-flavour scheme where 
the bottom-quark is treated as a massless particle and the bottom-Yukawa coupling is kept finite. 
We derived analytic expressions of the finite remainders up to two-loop level in perturbative QCD, taking into account the complete colour structure.
This is the first time analytic results are derived for a full colour two-loop five-point amplitude with an external mass.

The finite remainders are written in terms of one-mass pentagon functions and rational coefficients, which we reconstructed from multiple numerical evaluations over finite fields.
Our finite-field framework combines the Feynman diagram, four-dimensional projector and integral reduction methods, with optimised IBP relations obtained by means of
\textsc{NeatIBP}. 
We employed a number optimisations to reduce the complexity of the functional reconstruction of the rational coefficients, such as finding linear relations among them and ans\"atze, guessing the denominators using letters of
the alphabet, and performing on-the-fly univariate partial fraction decomposition. We observe that the analytic reconstruction of the subleading colour amplitudes
requires a significantly larger number of sample points than in the leading colour case, and that the memory usage and evaluation time per point per finite field
are also substantially higher.
We alleviated this by exploiting linear relations among the rational coefficients across
various subleading colour contributions.
We simplified the resulting analytic expressions of the rational coefficients through multivariate partial fraction decomposition with \textsc{pfd-parallel}.

We confirm the observations made in Ref.~\cite{Badger:2024sqv} regarding the analytic structure of the amplitudes: the non-planar letters $\sqrt{\Sigma_5^{(i)}}$ ($i=1,\ldots,6$) and the non-planar pentagon functions which diverge in the physical region are absent from the bare amplitudes, while the letter $\sqrt{\Delta}_5$ drops out of the finite remainders.
These non-trivial cancellations call for a theoretical explanation.

We implemented all finite remainders in a \textsc{C++} library and assess the numerical stability of the evaluation of the hard functions.
We find that the stability and evaluation time are suitable for immediate deployment in the computation of differential distributions at NNLO QCD accuracy.
The numerical implementation is provided as a \textsc{C++} package.

Our results, together with the implementation of the pentagon functions in the \textsc{PentagonFunctions++} library,
open the door to an efficient computation of the full-colour double-virtual corrections to $pp\to H+2b$~jets at NNLO QCD accuracy in the 5FS, as well as in an approximated 4FS where the leading $m_b$ contributions are included through massification.
In addition, also by employing massification procedure, our full-colour two-loop $pp\to b\bar{b}H$ amplitudes can be used to approximate full-colour two-loop $pp\to t\bar{t}H$ amplitudes in the high-energy region~\cite{Devoto:2024nhl}.

\section*{Acknowledgements}
We are indebted to Jakub Kry\'s for collaboration in the initial stages of this project.
H.B.H.\ has been supported by an appointment to the JRG Program at the APCTP through the Science and Technology Promotion Fund and Lottery Fund
of the Korean Government and by the Korean Local
Governments~--~Gyeongsangbuk-do Province and Pohang City.
S.B.~acknowledges funding from the Italian Ministry of
Universities and Research (MUR) through FARE grant R207777C4R and
through grant PRIN 2022BCXSW9. Y.Z.\ is supported from the NSF of China
through Grants No.\ 12075234, 12247103, and 12047502.
S.Z.~was supported by the European Union’s Horizon Europe research and innovation programme under the Marie
Skłodowska-Curie grant agreement No.~101105486, and by the Swiss National Science Foundation (SNSF) under the Ambizione grant No.~215960.
R.P.~would like to express his gratitude to the APCTP for the hospitality while parts of this project have been carried out.

\appendix

\section{Description of the ancillary files}
\label{app:ancillaryfiles}

In this Appendix, we describe the content of the \textsc{Mathematica} ancillary files, which can be downloaded from~\cite{zenodo}.
The description of the \textsc{C++} library, on the other hand, can be found in the \verb=anc_cpp/README= in the ancillary files~\cite{zenodo}.
The analytic expressions are given separately for the finite remainders and the $\eps$-pole terms that contain the UV and IR singularities. The $L$-loop finite remainders $F^{(L)}$
are presented in the form
\begin{equation}
\label{eq:finremsparse}
	F^{(L)} = \sum_{p=0}^{-2L+4} \eps^p \, \sum_{ij} r_{i}(\vec{y}) \, S_{pij} \, m_j(f) \,,
	\qquad L \in \{1,2\} \,,
\end{equation}
where $r_i$ is a set of independent rational coefficients collected across finite remainders with the same helicity configuration and loop order, $S$ is a sparse matrix of rational numbers,
and $m_j(f)$ are monomials built up of pentagon functions, transcendental constants, square roots, and power of $x_1$. The latter are included in the monomials to keep them dimensionless.
The rational coefficients, $r_{i}(\vec{y})$ in Eq.~\eqref{eq:finremsparse}, are expressed in terms of distinct polynomial factors $y_i(\vec{x})$, common for each set of $r_i$'s.
We recall that the analytic expressions are derived with $x_1=1$. While the $x_1$ dependence in the monomials $m_j(f)$ is explicitly reinstated in the provided expressions, that of the rational coefficients is recovered in the numerical evaluation scripts.

All analytical results derived in this paper are distributed in \textsc{Mathematica}-readable format.
In the following we give the translations of the symbols used in the ancillary files to the notation used in the paper.
\begin{itemize}
\item Partonic scattering processes contributing to $pp\to b\bar{b}H$:
\begin{align}
\label{eq:translateproc}
\begin{aligned}
	\texttt{2g2bH} &= \bbggH \,, \\
	\texttt{2q2bH} &= \bbqqH \,.
\end{aligned}
\end{align}
\item Colour factors in $\bbggH$:
\begin{align}
\label{eq:translatecolgg}
\begin{aligned}
	\texttt{T2341}  &= (t^{a_3}t^{a_4})_{i_2}^{\;\;\bar{i}_1} \,, \\
	\texttt{d12d34} &= \delta_{i_2}^{\;\;\bar{i}_1} \delta^{a_3 a_4} \,.
\end{aligned}
\end{align}
\item Colour factors in $\bbqqH$:
\begin{align}
\label{eq:translatecolqq}
\begin{aligned}
	\texttt{d14d23} &= \delta_{i_4}^{\;\;\bar{i}_1}  \delta_{i_2}^{\;\;\bar{i}_3}  \,,  \\
	\texttt{d12d34} &= \delta_{i_2}^{\;\;\bar{i}_1}  \delta_{i_4}^{\;\;\bar{i}_3}  \,.
\end{aligned}
\end{align}
\item Helicity configurations:
\begin{align}
\label{eq:translatehel}
\begin{aligned}
	\texttt{pppp} &= ++++ \,, \\
	\texttt{pppm} &= +++- \,, \\
	\texttt{ppmm} &= ++-- \,.
\end{aligned}
\end{align}
\item Dimensional regulator, momentum-twistor variables, Mandelstam invariants and other kinematical quantities needed to evaluate the amplitude:
\begin{align}
\label{eq:translatekinematics}
\begin{aligned}
    \texttt{eps} & = \eps \,, \\
	\texttt{ex[i]} &= x_i \quad \forall i = 1,\dots,6 \,, \\
	\texttt{s[i,j]} \; \mathrm{or} \; \texttt{sij} &= s_{ij} \,, \\
	\texttt{s[i,j,k]}  \; \mathrm{or} \; \texttt{sijk} &= s_{ijk} \,, \\
	\texttt{tr5} & = \tr_5 \,, \\
	\texttt{trp[i,j,k,l]} & = \tr_+(ijkl) \,, \\
	\texttt{trm[i,j,k,l]} & = \tr_-(ijkl) \,. \\
\end{aligned}
\end{align}
\item Pentagon functions ($\texttt{F[w,i]}$), transcendental constants and square roots (see Refs.~\cite{Abreu:2023rco,Chicherin:2021dyp} for a more detailed description of these objects): \begin{align}
\label{eq:translatepfuncs}
\begin{aligned}
	\texttt{re[3,1]} &= \zeta_3 \,, \\
	\texttt{im[1,1]} & = \i \pi \,, \\
	\texttt{sqrtDelta5} & = \sqrt{\Delta_5} \,, \\
	\texttt{sqrtG3[i]} & = \sqrt{\Delta_3^{(i)}} \,, \\
	\texttt{sqrtSigma5[i]} & = \sqrt{\Sigma_5^{(i)}} \,.
\end{aligned}
\end{align}
\item Renormalisation scale:
\begin{align}
\label{eq:translatemu}
\begin{aligned}
	\texttt{mu} &= \mu \,. \\ 
\end{aligned}
\end{align}

\end{itemize}

The content of the ancillary files is organised according to:
\begin{itemize}
\item scattering channel \verb=<proc>=  (\verb=2g2bH=, \verb=2q2bH=),
\item loop order \verb=<loop>= (\verb=1=, \verb=2=),
\item colour factor \verb=<col>=: (\verb=T2341= and  \verb=d12d34= for $\bbggH$, \verb=d14d23= and  \verb=d12d34= for $\bbqqH$),
\item powers of $n_f$ (\verb=<a>=) and $N_c$ (\verb=<b>=),\footnote{For $A^{(L)}_{2}$ we include the $1/N_c$ factor
	(see Eq.~\eqref{eq:bbqqHcolourdecomposition}) in the labelling of the $N_c$ powers in the ancillary files.}
\item helicity configuration \verb=<hel>=: (\verb=pppp=, \verb=pppm=, \verb=ppmm=).
\end{itemize}
We here list and describe all files we provided.
\begin{itemize}

	\item \verb=amplitudes_<proc>/Tree_<proc>_<col>_Nfp<a>_Ncp<b>_<hel>.m=:
		tree-level helicity amplitude.

	\item \verb=amplitudes_<proc>/FiniteRemainder_indep_coeffs_y_<loop>L_<proc>_<hel>.m=:
        independent rational coefficients $r_i$ in Eq.~\eqref{eq:finremsparse} as functions of common factors \verb=y[i]= ($y_i$).

	\item \verb=amplitudes_<proc>/FiniteRemainder_sm_<loop>L_<proc>_<col>_Nfp<a>_Ncp<b>_<hel>.m=:
		rational matrix $S$ in Eq.~\eqref{eq:finremsparse}, written in \textsc{Mathematica}'s \verb=SparseArray= format. 

	\item \verb=amplitudes_<proc>/ys_<loop>L_<proc>_<hel>.m=: replacement rules for the common polynomial factors \verb=y[i]=  ($y_i$) in 
		terms of momentum-twistor variables \verb=ex[i]=  ($x_i$).

	\item \verb=amplitudes_<proc>/FunctionBasis_<proc>_<loop>L.m=: monomial basis of pentagon functions, transcendental constants and square roots
		($m_j(f)$ in Eq.~\eqref{eq:finremsparse}).

	\item \verb=amplitudes_<proc>/mudep_<loop>L_<proc>_<col>_Nfp<a>_Ncp<b>_<hel>.m=: $\mu$-restoring term derived from 
		Eqs.~\eqref{eq:mu1L}~and~\eqref{eq:mu2L}.

	\item \verb=amplitudes_<proc>/Poles_<loop>L_<proc>_<col>_Nfp<a>_Ncp<b>_<hel>.m=: pole terms built out of UV and IR counterterms,
		which take the same form as the bare helicity amplitude in Eq.~\eqref{eq:amppfuncs} except that
		the rational coefficients are given separately for each helicity configuration. The pole terms are given in the following format,
        \begin{center}
        \verb={coefficientrules,{coefficients,monomials}}= \,,
        \end{center}
        where \verb=coefficientrules= is a list of replacement rules defining the independent rational coefficients \verb=f[i]= 
        in terms of the momentum-twistor variables \verb=ex[i]=,
        \verb=monomials= and \verb=coefficients= are the lists of pentagon-function monomials and
		 of their coefficients, which depend on both \verb=eps= ($\eps$) and the independent rational coefficients \verb=f[i]=.

	\item \verb=F_permutations_sm/=: a folder containing the pentagon functions' permutation rules. The permutation to the ordering $\texttt{<perm>}$ of the external momenta is given by the matrix stored in \verb=sm_F_perm_2l5p1m_<perm>.m= (in \textsc{Mathematica}'s \verb=SparseArray=
		format). Multiplying the latter by the array of pentagon-function monomials in
		\verb=F_monomials.m= gives the permutation of the pentagon functions listed in the variable \verb=listOfFunctions= in \verb=utilities.m=.

	\item \verb=Evaluate_BareAmplitudes_<proc>.wl=: script to evaluate numerically the bare helicity partial amplitudes in the independent
		helicity configurations for the $gg\to\bar{b}bH$ and $\bar{q}q \to\bar{b}bH$ scattering channels
		as defined in Eq.~\eqref{eq:bbHchannels}.

	\item \verb=Evaluate_HardFunctions_<proc>.wl=: script to evaluate numerically the hard functions for all partonic channels
		contributing to $pp\to b\bar{b}H$ as listed in Eq~\eqref{eq:bbHchannels}.

	\item \verb=utilities.m=: a collection of auxiliary functions needed for the numerical evaluation scripts.

	\item \verb=IR_pole_operators/Z<loop>_<proc>.m=: $\mathbf{Z}$ operator entering the IR-subtraction term, see
		Eqs.~\eqref{eq:finiteremainder} and \eqref{eq:Zexpansion}.

	\item \verb=IR_pole_operators/usage.wl=: a script to construct the IR-subtraction terms from $\mathbf{Z}$.

\end{itemize}

\bibliographystyle{JHEP}
\bibliography{bbH_2L}

\end{document}